\documentclass[a4paper,11pt]{article}
\usepackage{jheppub}
\usepackage{lineno}

\usepackage{tikz}
\usepackage{bm}
\usepackage{multirow}
\usepackage{makecell}
\usepackage{booktabs}

\allowdisplaybreaks[4]

\newcommand{\SKLP}{State Key Laboratory of Particle Detection and Electronics, University of Science and Technology of China, Hefei 230026, Anhui, People’s Republic of China}
\newcommand{\USTC}{Department of Modern Physics, University of Science and Technology of China, Hefei 230026, Anhui, People’s Republic of China}

\title{\boldmath Two-loop planar master integrals for NNLO QCD corrections to W-pair production in quark-antiquark annihilation}

\author[a,b]{Wen-Jie He,}
\author[a,b,c]{Ren-You Zhang,}
\author[a,b]{Liang Han,}
\author[a,b]{Yi Jiang,}
\author[a,b]{Zhe Li,}
\author[a,b]{Xiao-Feng Wang,}
\author[a,b]{Shu-Xiang Li,}
\author[a,b]{Pan-Feng Li,}
\author[d]{Qing-hai Wang\,}

\emailAdd{zhangry@ustc.edu.cn}

\affiliation[a]{\SKLP}
\affiliation[b]{\USTC}
\affiliation[c]{Anhui Center for Fundamental Sciences in Theoretical Physics, University of Science and Technology of China, Hefei 230026, Anhui, People’s Republic of China}
\affiliation[d]{Department of Physics, National University of Singapore, Singapore 117551, Singapore}

\abstract{
The planar two-loop scalar Feynman integrals contributing to the massive NNLO QCD corrections for $W$-boson pair production via quark-antiquark annihilation can be classified into three family branches, each of which is reduced to a distinct set of master integrals (MIs), totaling $27$, $45$ and $15$, respectively. These MIs are analytically calculated using the method of differential equations, with solutions expanded as Taylor series in the dimensional regulator $\epsilon$. For the first two family branches, the differential systems can be successfully transformed into canonical form by adopting appropriate bases of MIs. This enables the MIs of these family branches to be expressed either as Goncharov polylogarithms (GPLs) or as one-fold integrals over GPLs, up to $\mathcal{O}(\epsilon^4)$. In contrast, the differential system for the third family branch can only be cast into a form linear in $\epsilon$ due to the presence of elliptic integrals. The solution to this linear-form differential system is expressed in an iterated form owing to the strictly lower-triangular structure of the coefficient matrices at $\epsilon = 0$. Our analytic expressions for these MIs are verified with high accuracy against the numerical results from the \texttt{AMFlow} package.
}


\begin{document}
\maketitle
\flushbottom

\section{Introduction}
\label{sec:1}
\par
The production of vector-boson pairs at the Large Hadron Collider (LHC) plays a pivotal role in studying the electroweak (EW) gauge symmetry structure within the framework of the Standard Model (SM) \cite{ATLAS:2016nqi,ATLAS:2019rob,ATLAS:2017bbg,CMS:2020mxy}. By analyzing these production processes, precise predictions for some key EW observables, such as anomalous triple gauge-boson couplings, can be scrutinized through comparisons with experimental data from LHC proton-proton collisions at center-of-mass energies of 7, 8 and 13 TeV \cite{ATLAS:2012mec,CMS:2013ant,ATLAS:2016zwm,CMS:2015tmu,CMS:2020ezf,ATLAS:2017bbg,ATLAS:2018mxa}. Among all weak gauge-boson pair production processes, $W$-pair production contributes significantly to event data at high-energy colliders due to its relatively large cross section. Within the SM, $W$-boson pairs are produced primarily via three channels: the foremost quark-antiquark annihilation channel $q\bar{q} \rightarrow W^+W^-$ \cite{Brown:1978mq}; the loop-induced gluon-gluon fusion channel $gg \rightarrow W^+W^-$, which becomes significant at high energies due to large gluon luminosity \cite{Glover:1988fe}; and the Higgs boson channel $gg \rightarrow H \rightarrow W^+W^-$, which, though an order of magnitude smaller than the other channels, plays a pivotal role in the discovery of the Higgs boson \cite{CMS:2019ekd,ATLAS:2018xbv,ATLAS:2018jym}. Precise theoretical studies for $W$-pair production are indispensable not only for investigating the properties of the $W$ boson, such as its mass and trilinear gauge couplings, but also for searching for the Higgs boson in the $W^+W^-$ channel. Therefore, providing accurate theoretical predictions for $W$-pair production is of paramount importance to ensure alignment with the remarkable precision achieved in experimental measurements.

\par
To match the precision demanded by both current and forthcoming experimental analyses, it is essential to refine theoretical predictions by incorporating higher-order quantum chromodynamics (QCD) and electroweak (EW) corrections. The next-to-leading order (NLO) QCD and EW corrections to on-shell $W$-boson pair production at hadron colliders were first studied in refs. \cite{Ohnemus:1991kk,Bierweiler:2012kw}, while the subsequent leptonic decays of $W^{\pm}$ bosons and off-shell effects were investigated in refs. \cite{Campbell:1999ah,Biedermann:2016guo}. One of significant advances in precision studies of $W$ physics is the calculation of next-to-next-to-leading order (NNLO) QCD corrections to the inclusive production of on-shell $W$-boson pairs at hadron colliders \cite{Gehrmann:2014fva}. The hadronic production of $W$-boson pairs with leptonic decays of $W^{\pm}$ was analyzed at QCD NNLO \cite{Grazzini:2016ctr} by using the available two-loop helicity amplitudes for $W$-pair production \cite{Caola:2014iua,Gehrmann:2015ora}, accounting for spin correlations, off-shell effects and non-resonant contributions. Notably, considering the leptonic decays of $W$ bosons results in a reduction in the total cross section compared to the on-shell approximation. Furthermore, the contributions from the loop-induced gluon-gluon fusion channel were calculated up to QCD NLO \cite{Caola:2015rqy,Caola:2016trd} by leveraging the two-loop helicity amplitudes for $gg \rightarrow W^+W^-$ \cite{vonManteuffel:2015msa,Caola:2015ila}. The combination of NNLO QCD and NLO EW corrections to hadronic production of $W$-boson pairs, including photon-induced channels, has been studied in detail \cite{Grazzini:2019jkl}. Recently, the $\text{nN}^2\text{LO}$ corrections to $W^+W^-$ production at the LHC, namely, the full NNLO corrections supplemented by NLO corrections to the loop-induced channels, have been studied in ref. \cite{Grazzini:2020stb}, in which the quark-gluon partonic channels were considered for the first time.

\par
Two-loop master integrals (MIs) for vector-boson pair production at hadron colliders have been extensively studied over the past decade or so. The massless MIs involved in the NNLO QCD corrections to $q\bar{q} \rightarrow VV$, with two off-shell legs of equal invariant mass, were meticulously detailed in refs. \cite{Gehrmann:2013cxs,Gehrmann:2014bfa,Papadopoulos:2014lla}, while those with two distinct off-shell legs were presented in refs. \cite{Henn:2014lfa,Caola:2014lpa,Anastasiou:2014nha}. These massless two-loop MIs can be elegantly expressed in terms of Goncharov polylogarithms (GPLs), which are key mathematical constructs for a wide variety of Feynman integrals. These analytic studies of MIs have laid the foundation for the full calculation of vector-boson pair production in hadron collisions at QCD NNLO. However, the two-loop MIs with massive propagators for $q\bar{q} \rightarrow VV$ remain unresolved, with the primary challenge being the presence of an additional mass scale compared to the massless case. For massive two-loop MIs, more intricate classes of functions beyond the scope of GPLs are involved, such as elliptic integrals in many instances \cite{Bonciani:2016qxi, Adams:2018bsn, Adams:2018kez, Moriello:2019yhu, Badger:2021owl}.

\par
In this paper, we study the two-loop scalar Feynman integrals contributing to the massive NNLO QCD corrections for $W$-pair production via light quark-antiquark annihilation. Following the standard procedure, we begin by classifying these two-loop scalar integrals into $17$ topologies belonging to six integral families. Using an equivalence relation, to be defined later, we identify four top-level family branches: three planar and one non-planar. In this work, we focus exclusively on the planar branches. Using the integration-by-parts recurrence relations, the scalar Feynman integrals in each top-branch can be further reduced to a finite basis of independent integrals, termed master integrals. We derive the analytic expressions for these MIs using the method of differential equations. It turns out that the differential equation systems of two top-branches can be cast into canonical form, allowing the solutions to be expressed in terms of GPLs or one-fold integrals over GPLs, up to the fourth order in $\epsilon$. However, the differential equations for the MIs of the third top-branch can only be arranged in a form linear in $\epsilon$, due to the involvement of elliptic integrals, and the solution is expressed in an iterative form. All the analytic results are validated by comparison with the numerical results obtained using the auxiliary mass flow method, demonstrating a high degree of accuracy.

\par
The rest of this paper is organized as follows. In section \ref{sec:2}, we establish our notations and classify the scalar Feynman integrals involved in the massive NNLO QCD corrections to $W$-boson pair production in quark-antiquark annihilation. Section \ref{sec:3} is dedicated to the analytic calculation of the MIs by solving three differential systems, each tailored to a specific planar top-branch. A numerical verification of our analytic expressions for these MIs is provided in section \ref{sec:4}. Finally, a brief summary is given in section \ref{sec:5}.

\section{Notations and conventions}
\label{sec:2}
\par
In this paper, we study $W$-boson pair production in light quark-antiquark annihilation,
\begin{equation}
q(p_1) + \bar{q}(p_2) \longrightarrow W^+(p_3) + W^-(p_4)\,,
\qquad
(q = u, d, s, c)\,,
\end{equation}
where the initial-state light quarks are considered massless and the four on-shell external momenta $p_i~ (i = 1, ..., 4)$ are taken to be incoming. The amplitude for this $2 \rightarrow 2$ scattering process can be expressed in terms of the Mandelstam invariants,
\begin{equation}
s = (p_1 + p_2)^2\,,
\qquad
t = (p_2 + p_3)^2\,,
\qquad
u = (p_1 + p_3)^2\,,
\end{equation}
which satisfy $s + t + u = 2\, m_W^2$. All two-loop scalar Feynman integrals involved in the NNLO QCD corrections to this production channel can be classified into several integral families. In the dimensional regularization scheme, the dimensionless two-loop four-point scalar Feynman integrals of the family $\mathcal{F}( D_1, \ldots, D_9 )$ are conventionally defined as
\begin{equation}
\label{eq:FI}
F(n_1, \ldots, n_9)
=
\frac{1}{(Q^2)^{d - n}}
\int
\mathcal{D}^{d} l_1 \mathcal{D}^d l_2\,
\frac{1}{D_1^{n_1}  \ldots  D_9^{n_9}}\,,
\qquad
(n_i \in \mathbb{Z}, \quad i = 1, \ldots, 9)\,,
\end{equation}
where $n = n_1 + \cdots + n_9$, $d = 4 - 2\, \epsilon$ is the spacetime dimension, $Q$ is a characteristic mass scale, $\{ D_i \,|\, i = 1, \ldots, 9 \}$ is a complete set of independent propagators, $l_{1,2}$ are loop momenta, and the integration measure is taken as
\begin{equation}
\mathcal{D}^d l
=
\frac{d^d l}{(2 \pi)^d}
\left( \frac{i S_\epsilon}{16 \pi^2} \right)^{-1}
\quad
\text{with}
\quad
S_{\epsilon}
=
(4\pi)^{\epsilon}\, \Gamma(1+\epsilon)\,.
\end{equation}
Two Feynman integrals $F(n_1, \ldots, n_9)$ and $F(n_1^{\prime}, \ldots, n_9^{\prime})$ are considered equivalent if
\begin{equation}
\Theta(n_i - 1/2) = \Theta(n_i^{\prime} - 1/2)\,,
\qquad
(i = 1, \ldots, 9)\,.
\end{equation}
The equivalence classes induced by this equivalence relation, denoted as $[s_1, \ldots, s_9]$, where
\begin{equation}
[s_1, \ldots, s_9] = \{ F(n_1, \ldots, n_9) \,|\, \Theta(n_i - 1/2) = s_i \}\,,
\qquad
(s_i \in \{0, 1\}, \quad i = 1, \ldots, 9)\,,
\end{equation}
are called sectors or topologies and form a partition of the integral family. For two distinct sectors $\mathcal{S} = [s_1, \ldots, s_9]$ and $\mathcal{T} = [t_1, \ldots, t_9]$, where $s_i \leqslant t_i$ for $i \in \{1, \ldots, 9\}$, $\mathcal{S}$ is referred to as a sub-sector of $\mathcal{T}$ and $\mathcal{T}$ a super-sector of $\mathcal{S}$, denoted by
\begin{equation}
\mathcal{S} \prec \mathcal{T}
\quad
\text{and}
\quad
\mathcal{T} \succ \mathcal{S}\,.
\end{equation}
For a given sector $\mathcal{T}$, the family branch induced by $\mathcal{T}$ is defined as
\begin{equation}
\mathcal{T}_{\digamma} = \bigcup_{\scriptscriptstyle{\mathcal{S} \prec \mathcal{T}}} \mathcal{S} \cup \mathcal{T}\,.
\end{equation}

\par
\begin{figure}[htbp]
\centering
\includegraphics[width=0.3\textwidth]{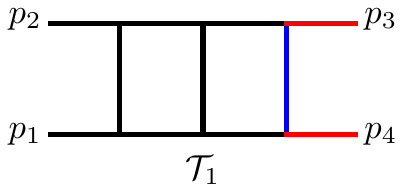}
\qquad
\includegraphics[width=0.3\textwidth]{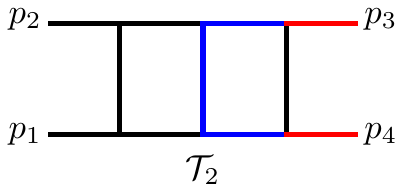}
\qquad
\includegraphics[width=0.3\textwidth]{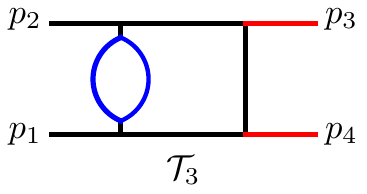}
\qquad
\includegraphics[width=0.3\textwidth]{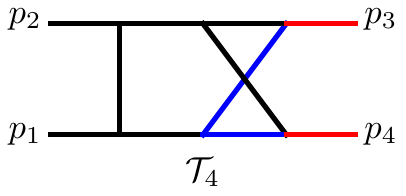}
\caption{
Top-sector diagrams for massive QCD two-loop corrections to $q\bar{q} \rightarrow W^+W^-$. The black, red and blue lines represent massless particles, $W$ bosons and top quarks, respectively.
}
\label{fig1}
\end{figure}

The massless two-loop four-point Feynman integrals with two off-shell legs of equal invariant mass, which are involved in the NNLO QCD corrections to $q\bar{q} \rightarrow VV$, have been thoroughly studied in refs. \cite{Gehrmann:2013cxs,Gehrmann:2014bfa}. Therefore, this paper will focus solely on the two-loop four-point integral functions with massive internal propagators. The irreducible QCD two-loop Feynman diagrams for $q\bar{q} \rightarrow W^+W^-$ with internal top-quark propagators can be classified into $17$ topologies, which belong to $6$ integral families.
\begin{itemize}
\item {\it Family} $\mathcal{F}_{\text{A}}$:
        \begin{equation}
        \begin{aligned}
        &
        D_1 = l_1^2
        &\quad&
        D_2 = (l_1 + p_1)^2
        &\quad&
        D_3 = (l_1 - p_2)^2
        \\
        &
        D_4 = l_2^2
        &\quad&
        D_5 = (l_2 + p_1)^2
        &\quad&
        D_6 = (l_2 - p_2)^2
        \\
        &
        D_7 = (l_1 - l_2)^2
        &\quad&
        D_8 = (l_1 - p_2 - p_3)^2
        &\quad&
        D_9 = (l_2 - p_2 - p_3)^2 - m_t^2
        \end{aligned}
        \end{equation}
        $2$ {\it topologies} $\subset \mathcal{F}_{\text{A}}$:
        \begin{equation}
        \mathcal{T}_{1} = [\,1,\, 1,\, 1,\, 0,\, 1,\, 1,\, 1,\, 0,\, 1\,]\,,
        \qquad
        \mathcal{T}_{1,1} = [\,0,\, 1,\, 1,\, 0,\, 1,\, 1,\, 1,\, 0,\, 1\,]
        ~~
        \end{equation}
\item {\it Family} $\mathcal{F}_{\text{B}}$:
        \begin{equation}
        \begin{aligned}
        &
        D_1 = l_1^2
        &\quad&
        D_2 = (l_1 + p_1)^2
        &\quad&
        D_3 = (l_1 - p_2)^2
        \\
        &
        D_4 = l_2^2 - m_t^2
        &\quad&
        D_5 = (l_2 + p_1)^2 - m_t^2
        &\quad&
        D_6 = (l_2 - p_2)^2 - m_t^2
        \\
        &
        D_7 = (l_1 - l_2)^2 - m_t^2
        &\quad&
        D_8 = (l_1 - p_2 - p_3)^2
        &\quad&
        D_9 = (l_2 - p_2 - p_3)^2
        \end{aligned}
        \end{equation}
        $6$ {\it topologies} $\subset \mathcal{F}_{\text{B}}$:
        \begin{equation}
        \mathcal{T}_{2} = [\,1,\, 1,\, 1,\, 0,\, 1,\, 1,\, 1,\, 0,\, 1\,]\,,
        \qquad
        \mathcal{T}_{2,1} = [\,0,\, 1,\, 1,\, 0,\, 1,\, 1,\, 1,\, 0,\, 1\,]
        \end{equation}
        \begin{equation}
        \mathcal{T}_{3} = [\,1,\, 1,\, 1,\, 1,\, 0,\, 0,\, 1,\, 1,\, 0\,]\,,
        \qquad
        \begin{aligned}
        &
        \mathcal{T}_{3,1} = [\,1,\, 1,\, 0,\, 1,\, 0,\, 0,\, 1,\, 1,\, 0\,]
        \\
        &
        \mathcal{T}_{3,2} = [\,1,\, 0,\, 1,\, 1,\, 0,\, 0,\, 1,\, 1,\, 0\,]
        \\
        &
        \mathcal{T}_{3,3} = [\,1,\, 0,\, 0,\, 1,\, 0,\, 0,\, 1, \,1,\, 0\,]
        \end{aligned}
        \end{equation}
\item {\it Family} $\mathcal{F}_{\text{C}}$:
        \begin{equation}
        \begin{aligned}
        &
        D_1 = l_1^2
        &\quad&
        D_2 = (l_1 + p_1)^2
        &\quad&
        D_3 = (l_1 - p_2)^2
        \\
        &
        D_4 = l_2^2
        &\quad&
        D_5 = (l_2 + p_1)^2
        &\quad&
        D_6 = (l_2 - p_2)^2
        \\
        &
        D_7 = (l_1 - l_2)^2
        &\quad&
        D_8 = (l_1 - l_2 - p_4)^2 - m_t^2
        &\quad&
        D_9 = (l_2 - p_2 - p_3)^2 - m_t^2
        \end{aligned}
        \end{equation}
        $2$ {\it topologies} $\subset \mathcal{F}_{\text{C}}$:
        \begin{equation}
        \mathcal{T}_{4} = [\,1,\, 1,\, 1,\, 0,\, 0,\, 1,\, 1,\, 1,\, 1\,]\,,
        \qquad
        \mathcal{T}_{4,1} = [\,0,\, 1,\, 1,\, 0,\, 0,\, 1,\, 1,\, 1,\, 1\,]
        \end{equation}
\item The remaining $3$ families and $7$ topologies can be derived from $\mathcal{F}_{\text{A}}$, $\mathcal{F}_{\text{B}}$, $\mathcal{F}_{\text{C}}$ and $\mathcal{T}_{1}$, $\mathcal{T}_{2}$, $\mathcal{T}_{3}$, $\mathcal{T}_{3,1}$, $\mathcal{T}_{3,2}$, $\mathcal{T}_{3,3}$, $\mathcal{T}_{4}$, respectively, by exchanging $p_1$ and $p_2$.\footnote{$\mathcal{T}_{1,1}$, $\mathcal{T}_{2,1}$ and $\mathcal{T}_{4,1}$ are invariant under the exchange of $p_1$ and $p_2$.}
\end{itemize}
The super-sub-sector relations of these topologies are shown diagrammatically as follows:
\begin{center}
\begin{tikzpicture}
\node[](text) at (-5,0){(top-sector)};
\node[](T1) at (-3,0) {$\mathcal{T}_{1}$};
\node[](T11) at (-3,-1.5) {$\mathcal{T}_{1,1}$};
\draw[line width=0.6pt](T1.south)--(T11.north);
\node[](T2) at (-1.5,0) {$\mathcal{T}_{2}$};
\node[](T21) at (-1.5,-1.5) {$\mathcal{T}_{2,1}$};
\draw[line width=0.6pt](T2.south)--(T21.north);
\node[](T3) at (0.75,0) {$\mathcal{T}_{3}$};
\node[](T31) at (0,-1.5) {$\mathcal{T}_{3,1}$};
\node[](T32) at (1.5,-1.5) {$\mathcal{T}_{3,2}$};
\node[](T33) at (0.75,-3) {$\mathcal{T}_{3,3}$};
\draw[line width=0.6pt](T3.south)--(T31.north);
\draw[line width=0.6pt](T3.south)--(T32.north);
\draw[line width=0.6pt](T31.south)--(T33.north);
\draw[line width=0.6pt](T32.south)--(T33.north);
\node[](T4) at (3,0) {$\mathcal{T}_{4}$};
\node[](T41) at (3,-1.5) {$\mathcal{T}_{4,1}$};
\draw[line width=0.6pt](T4.south)--(T41.north);
\node[](text) at (6,0){\qquad};
\end{tikzpicture}
\end{center}
The top-sectors $\mathcal{T}_{1}$, $\mathcal{T}_{2}$, $\mathcal{T}_{3}$ and $\mathcal{T}_{4}$ are depicted in figure \ref{fig1}. It is clear that $\mathcal{T}_{1}$, $\mathcal{T}_{2}$ and $\mathcal{T}_{3}$ are planar topologies, while $\mathcal{T}_{4}$ is non-planar. In this study, we focus primarily on planar topologies. All massive planar two-loop scalar Feynman integrals involved in the NNLO QCD corrections to $q\bar{q} \rightarrow W^+W^-$ belong to the top-branches $\mathcal{T}_{1\digamma}$, $\mathcal{T}_{2\digamma}$ and $\mathcal{T}_{3\digamma}$, i.e., the family branches induced by the top-sectors $\mathcal{T}_{1}$, $\mathcal{T}_{2}$ and $\mathcal{T}_{3}$, respectively.

\par
The integration-by-parts (IBP) recurrence relations \cite{Tkachov:1981wb,Chetyrkin:1981qh} show that scalar Feynman integrals within the same family are interrelated. After performing IBP reduction procedure, a finite basis of linearly independent integrals, known as master integrals, is obtained for a given integral set. Any dimensionally regularized scalar integral of this integral set can be expressed as a linear combination of the MIs, with coefficients that are rational functions of the kinematic variables and the dimensional regulator. In this work, we utilize \texttt{Kira} \cite{Maierhofer:2017gsa,Klappert:2020nbg}, based on Laporta’s algorithm \cite{Laporta:2000dsw}, to perform IBP reduction, yielding $27$, $45$ and $15$ MIs for the top-branches $\mathcal{T}_{1\digamma}$, $\mathcal{T}_{2\digamma}$ and $\mathcal{T}_{3\digamma}$, respectively.

\section{Differential equations}
\label{sec:3}
\par
The massive NNLO QCD corrections to $q\bar{q} \rightarrow W^+W^-$ encompass four distinct scales: $s$, $t$, $m_{\scriptscriptstyle{W}}^2$ and $m_t^2$. The characteristic mass scale $Q$, introduced for the nondimensionalization of Feynman integrals, is set to $m_t$ for convenience. Consequently, the dimensionless Feynman integrals defined in eq. \eqref{eq:FI} depend solely on the following three dimensionless ratios:
\begin{equation}
x = -\, \frac{s}{m_t^2}\,,
\qquad\quad
y = -\, \frac{t}{m_t^2}\,,
\qquad\quad
z = -\, \frac{m_{\scriptscriptstyle{W}}^2}{m_t^2}\,.
\end{equation}
Notably, the partial derivatives of the MIs of a given family branch with respect to kinematic variables can be expressed as linear combinations of the MIs themselves using IBP identities. In other words, the MIs within a family branch form a closed linear differential system. The method of differential equations \cite{Kotikov:1990kg,Gehrmann:1999as} offers several advantages for calculating MIs, leveraging potent mathematical techniques and systematic methodologies. In this section, we elaborate on the construction of differential equations for the MIs of the three planar top-branches $\mathcal{T}_{i \digamma}~ (i = 1, 2, 3)$ and present the solutions of these differential systems.

\par
The differential equations for a basis of MIs $\bm{f}(\bm{x}; \epsilon)$, derived using \texttt{LiteRed} \cite{Lee:2012cn,Lee:2013mka}, can be written as
\begin{equation}
\partial_{i} \bm{f}(\bm{x}; \epsilon)
=
\text{A}_{i}( \bm{x}; \epsilon)\, \bm{f}(\bm{x}; \epsilon)\,,
\end{equation}
where the coefficient matrices $\text{A}_{i}(\bm{x}; \epsilon)$ are composed of rational functions of $\bm{x}$ and $\epsilon$. However, the dependence of $\text{A}_{i}(\bm{x}; \epsilon)$ on $\epsilon$ is typically quite intricate, rendering analytical solutions challenging. A simplified approach was introduced in refs. \cite{Henn:2013pwa,Henn:2014qga}, proposing a linear transformation of the basis, denoted as $\bm{g}= \mathbb{T} \bm{f}$, to establish a new set of differential equations,
\begin{equation}
\label{eq:CDEs}
\partial_{i} \bm{g}(\bm{x}; \epsilon)
=
\epsilon\, \mathbb{A}_{i}(\bm{x})\,
\bm{g}(\bm{x}; \epsilon)\,,
\end{equation}
referred to as the canonical differential equations. The new coefficient matrices $\mathbb{A}_{i}(\bm{x})$ in these transformed differential equations are determined by
\begin{equation}
\epsilon\, \mathbb{A}_{i}
=
\mathbb{T}\, \text{A}_{i}\, \mathbb{T}^{-1} - \mathbb{T}\, \partial_{i} \mathbb{T}^{-1}\,.
\end{equation}
Various techniques and algorithms have been developed to construct the canonical differential equations \cite{Moser:1959,Lee:2014ioa,Gituliar:2017vzm,Magnus:1954zz,Argeri:2014qva,DiVita:2014pza,Hoschele:2014qsa,Dlapa:2020cwj,Henn:2022vqp,Dlapa:2021qsl,Henn:2020lye,Ma:2021cxg,Dlapa:2022wdu,Chen:2022lzr}. To facilitate a clearer understanding of the solution, the canonical differential system \eqref{eq:CDEs} is presented in the following $d\log$ form,
\begin{equation}
\label{eq:dlogTDE}
d \bm{g}(\bm{x}; \epsilon) = \epsilon\, d \mathbb{A}(\bm{x})\, \bm{g}(\bm{x}; \epsilon)\,,
\end{equation}
where the coefficient matrix $d \mathbb{A}(\bm{x})$ is a sum of $d\log$'s multiplied by constant matrices,
\begin{equation}
\label{eq:dlogMatrix}
d \mathbb{A}(\bm{x}) = \sum_{i=1}^{k}\, \mathbb{C}_{i}\, d \log \omega_{i}(\bm{x})\,.
\end{equation}
The general solution to all orders in $\epsilon$ can be readily expressed in terms of Chen's iterated integrals \cite{Chen:1977oja},
\begin{equation}
\bm{g}(\bm{x}; \epsilon)
=
\mathcal{P} \exp \Big( \epsilon \int_{\gamma} d \mathbb{A} \Big)\, \bm{g}(\bm{x}_0; \epsilon)\,,
\end{equation}
where $\mathcal{P}$ denotes the path ordering of the matrix exponential, $\bm{g}(\bm{x}_{0}; \epsilon)$ represents the boundary values of the MIs, and the integration path $\gamma$ connects $\bm{x}_{0}$ to $\bm{x}$ within the space of kinematic variables. When all symbol letters $\omega_{1, \ldots, k}(\bm{x})$ are rational functions, the solution can be directly expressed in terms of Goncharov polylogarithms, which are defined recursively by
\begin{equation}
G(a_n, \ldots, a_1; z)
=
\int_{0}^{z}{\frac{dz^{\prime}}{z^{\prime} - a_n}G(a_{n-1}, \ldots, a_1; z^{\prime})}
\end{equation}
with $G(\,;z)=1$ and
\begin{equation}
G(\underbrace{0, \ldots, 0}_{n-\text{times}}; z)
=
\frac{\log^n(z)}{n!}\,,
\end{equation}
where $(a_n, \ldots, a_1)$ is referred to as the weight vector of the weight-$n$ GPL $G(a_n, \ldots, a_1; z)$.

\par
For the differential systems of the first two top-branches, $\mathcal{T}_{1\digamma}$ and $\mathcal{T}_{2\digamma}$, the solutions can be elegantly expressed in terms of GPLs or iterated integrals over GPLs. However, for the top-branch $\mathcal{T}_{3\digamma}$, where elliptic Feynman integrals are involved, achieving a canonical form for the differential system becomes elusive. Nevertheless, we can construct a set of ``relaxed'' canonical differential equations of the form\footnote{The differential equations \eqref{eq:LDEs} are also known as linear-form differential equations, and thus $\bm{g}(\bm{x}; \epsilon)$ is conventionally referred to as a linear basis of MIs.}
\begin{equation}
\label{eq:LDEs}
\partial_{i} \bm{g}(\bm{x}; \epsilon)
=
\Big[\,
\mathbb{A}_{i}^{(0)}(\bm{x}) + \epsilon\, \mathbb{A}_{i}^{(1)}(\bm{x})
\,\Big]\,
\bm{g}(\bm{x}; \epsilon)\,,
\end{equation}
where $\mathbb{A}_{i}^{(0)}(\bm{x})$ are strictly lower-triangular matrices \cite{Adams:2018kez,Badger:2021owl,Adams:2018bsn}. Consequently, the solution for the top-branch $\mathcal{T}_{3\digamma}$ can be expressed in an iterated form.

\subsection{Top-branch $\mathcal{T}_{1\digamma}$}
\label{subsec:3.1}
\par
\begin{figure}[htbp]
\centering
\includegraphics[width=1\textwidth]{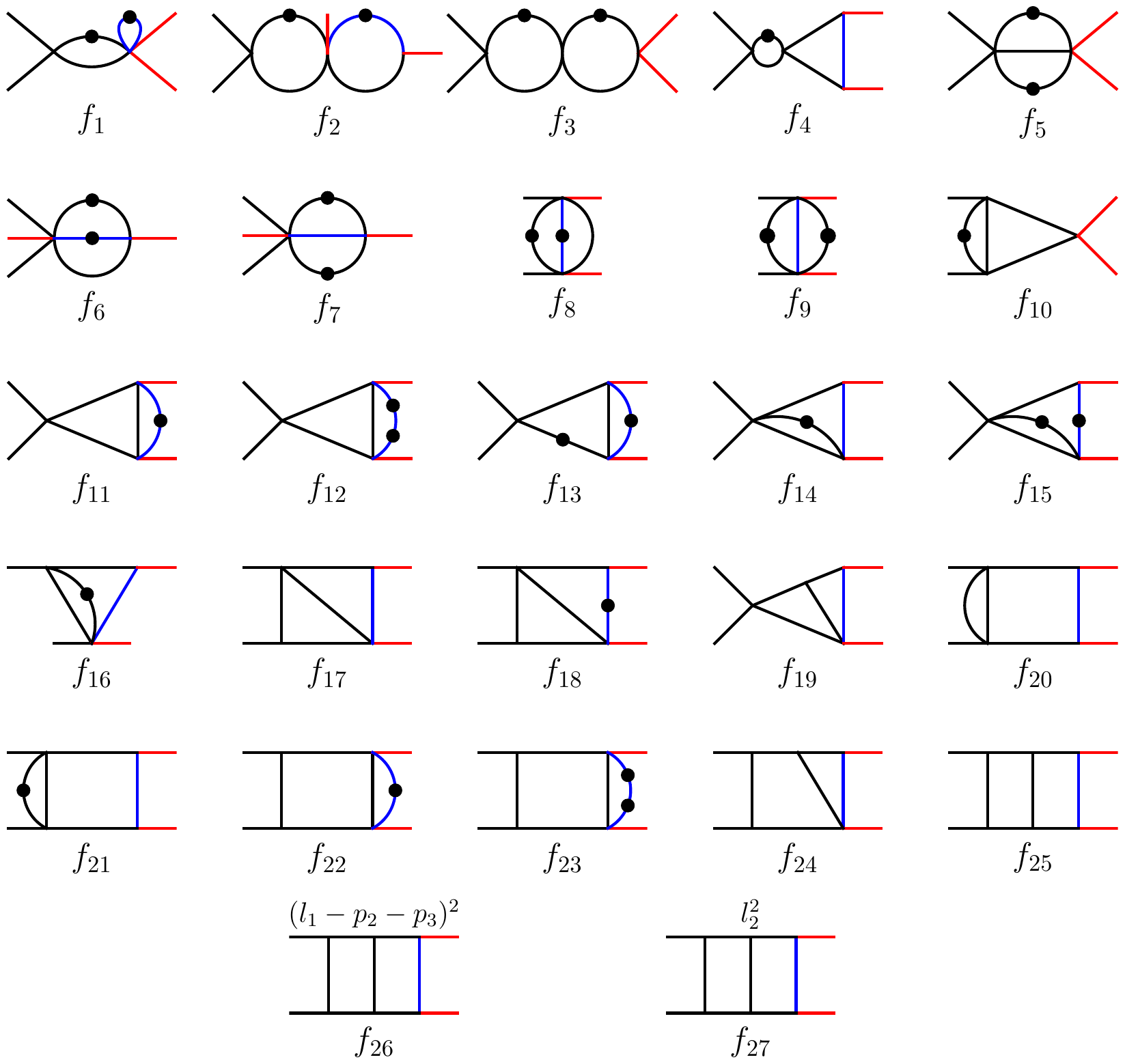}
\caption{
A basis of MIs for the top-branch $\mathcal{T}_{1\digamma}$. The dots denote additional powers of the corresponding propagators.
}
\label{fig2}
\end{figure}

With the help of \texttt{Kira}, we obtain a set of $27$ MIs of the top-branch $\mathcal{T}_{1\digamma}$, as shown in figure \ref{fig2}. The canonical basis of MIs, $\bm{g} = (g_1, \ldots, g_{27})^{T}$, is defined by the following transformation:
\begin{align}
\label{eq:gT1}
g_{1} & = \epsilon^2\, f_{1}\, x\,,  &
g_{2} & = \epsilon^2\, f_{2}\, x\, z\,,  &
\nonumber \\
g_{3} & = \epsilon^2\, f_{3}\, x^2\,,  &
g_{4} & = \epsilon^3\, f_{4}\, x\, r\,,  &
\nonumber \\
g_{5} & = \epsilon^2\, f_{5}\, x\,,  &
g_{6} & = \epsilon^2\, f_{6}\, z\,,  &
\nonumber \\
g_{7} & = \epsilon^2\, f_{7}\, (z + 1) + 2\, \epsilon^2\, f_{6}\,,  &
g_{8} & = \epsilon^2\, f_{8}\, y\,,  &
\nonumber \\
g_{9} & = \epsilon^2\, f_{9}\, (y + 1) + 2\, \epsilon^2\, f_{8}\,,  &
g_{10} & = \epsilon^3\, f_{10}\, x\,,  &
\nonumber \\
g_{11} & = \epsilon^3\, f_{11}\, r\,,  &
g_{12} & = \epsilon^2\, f_{12}\, r\,,  &
\nonumber \\
g_{13} & = \epsilon^2\, f_{13}\, x \, z
               + \epsilon^2\, f_{12}\, x
               + 3/2\, \epsilon^3\, f_{11}\, x\,,  &
g_{14} & = \epsilon^3\, f_{14}\, r\,,  &
\\
g_{15} & = \epsilon^2\, f_{15}\, [\, x - (z + 1)^2 \,]
               + 3/2\, \epsilon^3\, f_{14}\, [\, x - 2\, (z + 1) \,]\,,  &
g_{16} & = \epsilon^3\, f_{16}\, (y - z)\,,  &
\nonumber \\
g_{17} & = \epsilon^4\, f_{17}\, (x + y - z)\,,  &
g_{18} & = \epsilon^3\, f_{18}\, [\, x + (z + 1)\, (y - z) \,]\,,  &
\nonumber \\
g_{19} & = \epsilon^4\, f_{19}\, r\,,  &
g_{20} & = \epsilon^3\, (1- 2\, \epsilon)\, f_{20}\, r\,,  &
\nonumber \\
g_{21} & = \epsilon^3\, f_{21}\, x\, (y + 1)\,,  &
g_{22} & = \epsilon^3\, f_{22}\, x\, y\,,  &
\nonumber \\
g_{23} & = \epsilon^2\, f_{23}\, x\, (y + 1)
               + \epsilon^3\, f_{22}\, x\,,  &
g_{24} & = \epsilon^4\, f_{24}\, x\, (y - z)\,,  &
\nonumber \\
g_{25} & = \epsilon^4\, f_{25}\, x^2\, (y + 1)\,,  &
g_{26} & = \epsilon^4\, f_{26}\, x\, r\,,  &
\nonumber \\
g_{27} & = (y + 1)^{-1} \sum_{i}\, \alpha_{i}\, f_{i}\,,  &
\nonumber
\end{align}
where $r$ is the square root of $x\,(x - 4\, z)$, and the coefficients $\alpha_i$ in $g_{27}$ are provided in appendix \ref{appendix:A}. This canonical basis of MIs satisfies the $d\log$-form total differential equation \eqref{eq:dlogTDE}. Furthermore, following eq. \eqref{eq:dlogMatrix}, the $14$ symbol letters are as follows:
\begin{align}
\omega_{1} & = x\,,  &
\omega_{2} & = y\,,  &
\omega_{3} & = z\,,  &
\nonumber \\
\omega_{4} & = z + 1\,,  &
\omega_{5} & = y + 1\,,  &
\omega_{6} & = z - y\,,  &
\nonumber \\
\omega_{7} & = x - 4\, z\,,  &
\omega_{8} & = x + y - z\,,  &
\omega_{9} & = x - (z + 1)^2\,,  &
\\
\omega_{10} & = x + (z + 1)\, (y - z)\,,  &
\omega_{11} & = x\, y + (y - z)^2\,,  &
\omega_{12} & = (x - r)/(x + r)\,,  &
\nonumber \\
\omega_{13} & = \frac{x - 2\, (z + 1) - r}{x - 2\, (z + 1) + r}\,,  &
\omega_{14} & = \rlap{$\displaystyle \frac{x + 2\, (y - z) - r}{x + 2\, (y - z) + r}\,.$}
\nonumber
\end{align}
The square root $r$ can be rationalized by the following change of variables,
\begin{equation}
x = \frac{(x_1 + z)^2}{x_1}\,.
\end{equation}
Consequently, all the letters are rational functions of the variables $x_1$, $y$ and $z$. Thus, we can take advantage of the GPLs to construct the solution of this canonical differential system, leaving only the integration constants undetermined. 

\par
To arrive at a definite solution for a differential system of MIs, the integration constants{\textemdash}namely, the boundary conditions of the MIs{\textemdash}must be specified. Certain MIs, for which analytic expressions are available in existing literature, can be treated as independent inputs. For other MIs, the regularity properties at spurious singularities are used to determine the integration constants. This process imposes constraints on the integration constants, and the attainment of an exact solution hinges on establishing a sufficient number of these constraints. The regularity properties of Feynman integrals can be analyzed using the expansion by regions technique \cite{Beneke:1997zp,Smirnov:1994tg}, with the aid of the \texttt{asy.m} code \cite{Pak:2010pt,Jantzen:2011nz,Jantzen:2012mw}, which is available in the \texttt{FIESTA} package \cite{Smirnov:2015mct,Smirnov:2021rhf}. For the differential system of $\mathcal{T}_{1\digamma}$, the boundary conditions imposed on the MIs to determine the integration constants are listed below.
\begin{itemize}
\item $g_{1, 3, 5, \ldots, 10}$ are taken from refs. \cite{DiVita:2017xlr,Long:2021fdc} as independent inputs.
\item The integration constants of the remaining $19$ MIs are specified by the following regularity conditions:
      \begin{itemize}
      \item regularity at $x = 0$: $g_{14, 19, 26}$
      \item regularity at $y = 0$: $g_{22}$
      \item regularity at $z = 0$: $g_{2, 11, 14, 16, 17, 20, 23, \ldots, 27}$
      \item regularity at $x/z = 4$: $g_{4, 11, 12, 20}$
      \end{itemize}
\end{itemize}
In appendix \ref{appendix:B1}, we showcase the explicit expressions for the canonical MIs $g_i~ (i = 1, \ldots, 27)$ of the top-branch $\mathcal{T}_{1\digamma}$ up to $\mathcal{O}(\epsilon^2)$. The analytic expressions up to $\mathcal{O}(\epsilon^4)$ are available in the supplementary file ``analytic\_T1.m.''

\subsection{Top-branch $\mathcal{T}_{2\digamma}$}
\label{subsec:3.2}
\par
\begin{figure}[htbp]
\centering
\includegraphics[width=1\textwidth]{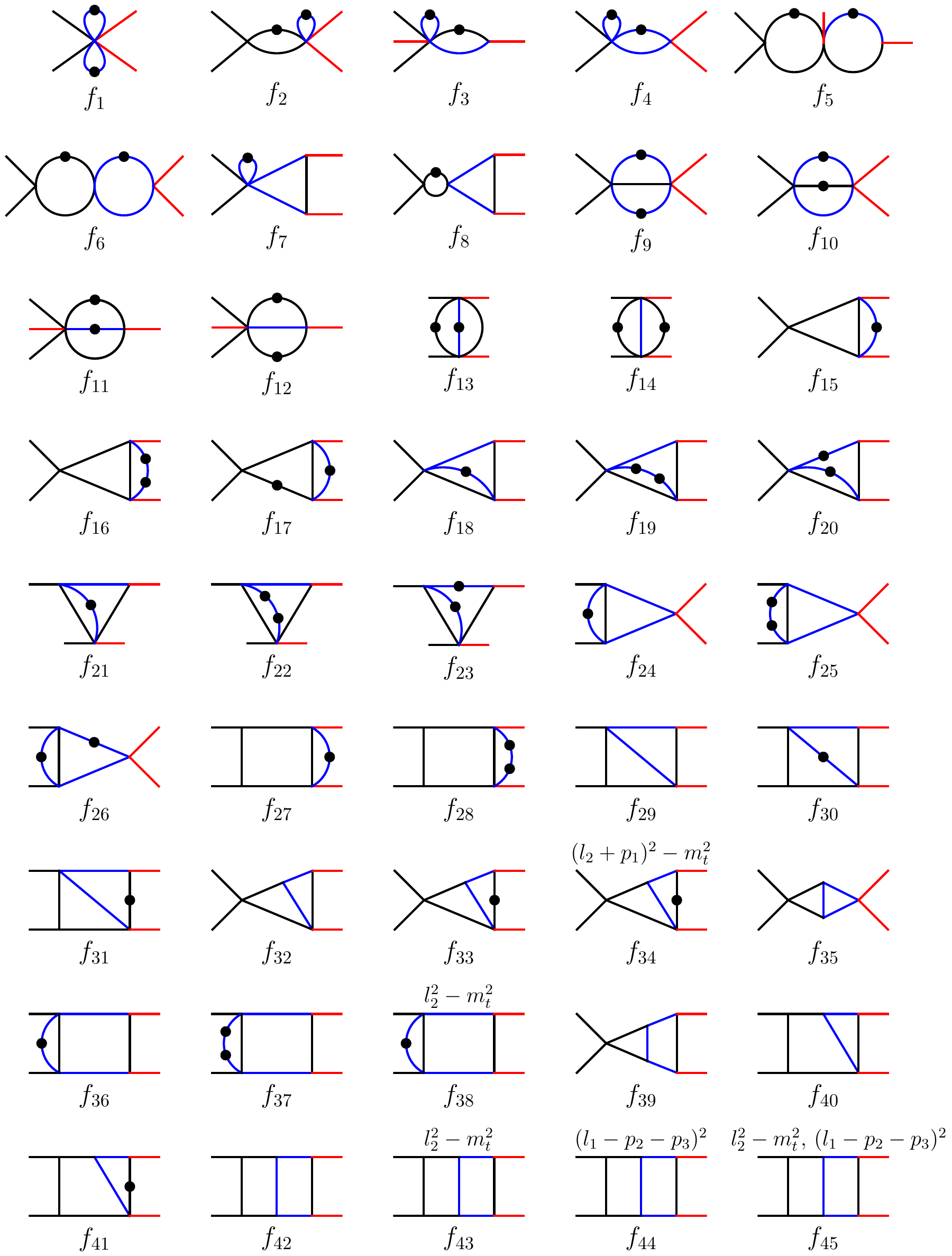}
\caption{
A basis of MIs for the top-branch $\mathcal{T}_{2\digamma}$.
}
\label{fig3}
\end{figure}

A pre-canonical basis of MIs for the top-branch $\mathcal{T}_{2\digamma}$, denoted by $\bm{f} = (f_{1}, \ldots, f_{45})^{T}$, is illustrated in figure \ref{fig3}. The linear transformation from the pre-canonical basis $\bm{f}$ to the canonical basis $\bm{g}$ is given as follows:
\begin{align}
\label{eq:gT2}
g_{1} & = \epsilon^2\, f_{1}\,,  &
g_{2} & = \epsilon^2\, f_{2}\, x\,,  &
\nonumber \\
g_{3} & = \epsilon^2\, f_{3}\, z\,,  &
g_{4} & = \epsilon^2\, f_{4}\, r_1\,,  &
\nonumber \\
g_{5} & = \epsilon^2\, f_{5}\, x\, z\,,  &
g_{6} & = \epsilon^2\, f_{6}\, x\, r_1\,,  &
\nonumber \\
g_{7} & = \epsilon^3\, f_{7}\, r_2\,,  &
g_{8} & = \epsilon^3\, f_{8}\, x\, r_2\,,  &
\nonumber \\
g_{9} & = \epsilon^2\, f_{9}\, x\,,  &
g_{10} & = \epsilon^2\, f_{10}\, r_1 + 1/2\, \epsilon^2\, f_9\, r_1\,,  &
\nonumber \\
g_{11} & = \epsilon^2\, f_{11}\, z\,,  &
g_{12} & = \epsilon^2\, f_{12}\, (z + 1)
               + 2\, \epsilon^2\, f_{11}\,,  &
\nonumber \\
g_{13} & = \epsilon^2\, f_{13}\, y\,,  &
g_{14} & = \epsilon^2\, f_{14}\, (y + 1)
               + 2\, \epsilon^2\, f_{13}\,,  &
\nonumber \\
g_{15} & = \epsilon^3\, f_{15}\, r_2\,,  &
g_{16} & = \epsilon^2\, f_{16}\, r_2\,,  &
\nonumber \\
g_{17} & = \epsilon^2\, f_{17}\, x\, z
               + \epsilon^2\, f_{16}\, x
               + 3/2\, \epsilon^3\, f_{15}\, x\,,  &
g_{18} & = \epsilon^3\, f_{18}\, r_2\,,
\nonumber \\
g_{19} & = \rlap{$\displaystyle
                  \epsilon^2\, f_{19}\, r_2\,,
\qquad\qquad
g_{20} = \epsilon^2\, f_{20}\, [\, 1 + x\, (z + 1) \,]
               + \epsilon^2\, f_{19}\, (x + 2)
               + 3/2\, \epsilon^3\, f_{18}\, (x + 2)\,,$}  &
\nonumber \\
g_{21} & = \epsilon^3\, f_{21}\, (y - z)\,,  &
g_{22} & = \epsilon^2\, f_{22}\, (y - z)\,,  &
\nonumber \\
g_{23} & = \epsilon^2\, f_{23}\, (1 - y + z)
               + 2\, \epsilon^2\, f_{22}
               + 3\, \epsilon^3\, f_{21}\,,  &
g_{24} & = \epsilon^3\, f_{24}\, x\,,  &
\nonumber \\
g_{25} & = \epsilon^2\, f_{25}\, x\,,  &
g_{26} & = \epsilon^2\, f_{26}\, r_1
               + \epsilon^2\, f_{25}\, r_1
               + 3/2\, \epsilon^3\, f_{24}\, r_1\,,  &
\nonumber \\
g_{27} & = \epsilon^3\, f_{27}\, x\, y\,,  &
g_{28} & = \epsilon^2\, f_{28}\, x\, (y + 1)
               + \epsilon^3\, f_{27}\, x\,,  &
\nonumber \\
g_{29} & = \epsilon^4\, f_{29}\, (x + y - z)\,,  &
g_{30} & = \epsilon^3\, f_{30}\, r_3\,,  &
\\
g_{32} & = \epsilon^4\, f_{32}\, r_2\,,  &
g_{33} & = \epsilon^3\, f_{33}\, (z + 1)\, r_2\,,  &
\nonumber \\
g_{34} & = \rlap{$\displaystyle
                  \epsilon^3\, f_{34}\, z
               + 1/2\, \epsilon^3\, f_{33}\, x\, (z + 1)
                - \epsilon^4\, f_{32}\, x
               + \epsilon^2\, f_{16}\, (x - 2\, z)
               +1/2\, \epsilon^3\, f_{15}\, (x - 2\, z)\,,$}  &
\nonumber \\
g_{35} & = \epsilon^3\, (1- 2\, \epsilon)\, f_{35}\, x\,,  &
g_{36} & = \epsilon^3\, f_{36}\, r_4\,,  &
\nonumber \\
g_{37} & = \epsilon^2\, f_{37}\, r_3 + \epsilon^3\, f_{36}\, r_3\,,  &
g_{38} & = \epsilon^3\, f_{38}\, r_2 + \epsilon^3\, f_{36}\, r_2\,,  &
\nonumber \\
g_{39} & = \epsilon^4\, f_{39}\, r_1\, r_2\,,  &
g_{40} & = \epsilon^4\, f_{40}\, x\, (y - z)\,,  &
\nonumber \\
g_{41} & = \rlap{$\displaystyle
              [\, \epsilon^3\, f_{41}\, (y - z)
               + \epsilon^3\, f_{30}\,
               - 2\, \epsilon^2\, f_{28}\,
               - 2\, \epsilon^3\, f_{27} \,]\, x\, (z + 1)\,,$}  &
\nonumber \\
g_{42} & = \epsilon^4\, f_{42}\, x\, r_3\,,  &
g_{43} & = \epsilon^4\, f_{43}\, x\, r_2\,,  &
\nonumber \\
g_{44} & = \rlap{$\displaystyle
                  \epsilon^4\, f_{44}\, x\, r_1
               + \epsilon^4\, f_{42}\, x\, y\, r_1
               - 2\, \epsilon^2\, f_{37}\, (y - z)\, r_1
               - 2\, \epsilon^3\, f_{36}\, (y - z)\, r_1
               + \epsilon^3\, f_{30}\, (y - z)\, r_1\,,$}  &
\nonumber \\
g_{31} & = \big[\, x\, (z + 1) \,\big]^{-1} \sum_{i}\, \beta_{i}\, f_{i}\,,  &
g_{45} & = \sum_{i}\, \gamma_{i}\, f_{i}\,,  &
\nonumber
\end{align}
where the four square roots $r_{1, 2, 3, 4}$ are defined by
\begin{align}
r_1^2 & = x\, (x + 4)\,,  &
r_2^2 & = x\, (x - 4\, z)\,,
\nonumber \\
r_3^2 & = x\, \big[\, x\, (y + 1)^2 + 4\, (y - z)^2 \,\big]\,,  &
r_4^2 & = x\, y\, \big[\, 4\, (z + 1) - y\, (x + 4) \,\big]\,,
\end{align}
and the coefficients $\beta_i$ in $g_{31}$ and $\gamma_i$ in $g_{45}$ are presented in appendix \ref{appendix:A}. This set of $45$ canonical MIs satisfies the $d\log$-form total differential equation \eqref{eq:dlogTDE}, with the corresponding $35$ symbol letters listed below:
\begin{align}
\omega_{1} & = x\,,  &
\omega_{2} & = y\,,  &
\omega_{3} & = z\,,  &
\nonumber \\
\omega_{4} & = x + 4\,,  &
\omega_{5} & = y + 1\,,  &
\omega_{6} & = z + 1\,,  &
\nonumber \\
\omega_{7} & = x - 4\, z\,,  &
\omega_{8} & = z - y\,,  &
\omega_{9} & = 1 - y + z\,,  &
\nonumber \\
\omega_{10} & = x + y - z\,,  &
\omega_{11} & = x\, (z + 1) + 1\,,  &
\omega_{12} & = x\, (z + 1) + (y - z)\,,  &
\nonumber \\
\omega_{13} & = x - (z + 1)^2\,,  &
\omega_{14} & = x\, y + (y - z)^2\,,  &
\omega_{15} & = x\, (y + 1)^2 + 4\, (y - z)^2\,,  &
\nonumber \\
\omega_{16} & = \frac{x - r_1}{x + r_1}\,,  &
\omega_{17} & = \frac{x - r_2}{x + r_2}\,,  &
\omega_{18} & = \frac{x + 2 - r_2}{x + 2 + r_2}\,,  &
\nonumber \\
\omega_{19} & = \frac{x - 2\, (z + 1) - r_2}{x - 2\, (z + 1) + r_2}\,,  &
\omega_{20} & = \frac{x + 2\, (y - z) - r_2}{x + 2\, (y - z) + r_2}\,,  &
\omega_{21} & = \frac{x - x\, y - r_3}{x - x\, y + r_3}\,,  &
\nonumber \\
\omega_{22} & = \frac{x + x\, y - r _3}{x + x\, y + r _3}\,,  &
\omega_{23} & = \frac{x - x\, y + 2\, x\, z - r_3}{x - x\, y + 2\, x\, z + r_3}\,,  &
\omega_{24} & = \frac{x\, y - r_4}{x\, y + r_4}\,,  &
\\
\omega_{25} & = \frac{x\, y + 2\, (y - z) - r_4}{x\, y + 2\, (y - z) + r_4}\,,  &
\omega_{26} & = \frac{r_1 - r_2}{r_1 + r_2}\,,  &
\omega_{27} & = \frac{y\, r_1 - r_4}{y\, r_1 + r_4}\,,  &
\nonumber \\
\omega_{28} & = \frac{y\, r_2 - r_4}{y\, r_2 + r_4}\,,  &
\omega_{29} & = \frac{(y + 1)\, r_1 - r_3}{(y + 1)\, r_1 + r_3}\,,  &
\omega_{30} & = \frac{(y + 1)\, r_2 - r_3}{(y + 1)\, r_2 + r_3}\,,  &
\nonumber \\
\omega_{31} & = \frac{(y - z)\, r_1 - r_3}{(y - z)\, r_1 + r _3}\,,  &
\omega_{32} & = \frac{y\, r_3 - (y + 1)\, r_4}{y\, r_3 + (y + 1)\, r_4}\,,  &
\nonumber \\
\omega_{33} & = \rlap{$\displaystyle
                            \frac{(1 - y)\, (z - y)\, r_2 - (y + z)\, r_3}{(1 - y)\, (z - y)\, r_2 + (y + z)\, r_3}\,,
                            \qquad\qquad
\omega_{34}    = \frac{y\, (2 - y + z)\, r_2 - (y + z)\, r_4}{y\, (2 - y + z)\, r_2 + (y + z)\, r_4}\,,$}  &
\nonumber \\
\omega_{35} & = \rlap{$\displaystyle
                            \frac{y\, (2 - y + z)\, r_3 - (1 - y)\, (y - z)\, r_4}{y\, (2 - y + z)\, r_3 + (1 - y)\, (y - z)\, r_4}\,.$}  &
\nonumber
\end{align}

\par
Compared to the differential system of the top-branch $\mathcal{T}_{1\digamma}$, that of $\mathcal{T}_{2\digamma}$ is more intricate due to the increased number of square roots involved. Achieving a simultaneous rationalization of the four square roots $r_i~ (i = 1, \ldots, 4)$ proves elusive. However, with the assistance of the \texttt{RationalizeRoots} package \cite{Besier:2019kco}, we can simultaneously rationalize $r_1$, $r_2$ and $r_3$ using the following change of variables, 
\begin{equation}
\label{eq:XYZchange}
(x, y, z)\, \longmapsto \,(x_2, y_2, z_2):
\quad
\left\lbrace\,
\begin{aligned}
&
x = \frac{x_2^2}{x_2 + 1}
\\
&
y =
\frac{x_2\, (x_2 + 1)\, y_2\, z_2^2 - x_2^2\, \big[\, y_2^2 - (x_2 +1) \,\big]\, (z_2 + 1)}{(x_2 + 1)\, (y_2 + 1)\, \big[\, y_2 - (x_2 + 1) \,\big]\, z_2^2}
\\
&
z = -\, \frac{x_2^2\, (z_2 + 1)}{z_2^2\, (x_2 + 1)}
\end{aligned}
\right.
\end{equation}
Consequently, the entire differential system is left with a single unrationalizable square root, which can be conveniently chosen as
\begin{equation}
R
=
\sqrt{
\big[\, \lambda_1\, z_2^2 + \lambda_2\, (z_2 + 1) \,\big]
\,
\big[\, \lambda_3\, z_2^2 + \lambda_4\, (z_2 + 1) \,\big]
}\,,
\end{equation}
where $\lambda_i~ (i = 1, \ldots, 4)$ are polynomials in $x_2$ and $y_2$, expressed in the following form:
\begin{equation}
\lambda_i = \mu_i\, y_2 + \nu_i\, \big[\, y_2^2 - (x_2 + 1) \,\big]\,,
\end{equation}
with
\begin{equation}
\begin{aligned}
&
\mu_1 = x_2\, (x_2 + 1)\,,
&\qquad\quad&
\nu_1 = 0\,,
\\
&
\mu_2 = 0\,,
&\qquad\quad&
\nu_2 = -\, x_2^2\,,
\\
&
\mu_3 = x_2\, (x_2 + 1)\, \big[\, 8\, (x_2 + 1) + x_2^2 \,\big]\,,
&\qquad\quad&
\nu_3 = -\, 4\, (x_2 + 1)^2\,,
\\
&
\mu_4 = -\, 4\, x_2^3\, (x_2 + 1)\,,
&\qquad\quad&
\nu_4 = -\, x_2^4\,.
\end{aligned}
\end{equation}
The presence of this unrationalizable square root makes it impossible to express all $45$ canonical MIs of $\mathcal{T}_{2\digamma}$ in terms of conventional GPLs. These canonical MIs can be expressed collectively as a path-ordered integral. It is essential to choose an appropriate initial point for the integration path, at which the values of these MIs can be easily obtained.

\par
For the top-branch $\mathcal{T}_{2\digamma}$, the coefficient matrix $\mathbb{A}_{x}$ can be expanded into a Taylor series in terms of the small variables $y$ and $z$, while $\mathbb{A}_{y}$ and $\mathbb{A}_{z}$ are expanded in Laurent series,
\begin{equation}
\left\lbrace~
\begin{aligned}
&
\mathbb{A}_{x}(x, y, z) = \sum_{m, n = 0}^{+\infty} \mathbb{A}_{x, (m,n)}(x)\, y^m z^n\,,
&
\\
&
\mathbb{A}_{y}(x, y, z) = \frac{\mathbb{A}_{y, (-1,0)}}{y} + \sum_{m, n = 0}^{+\infty} \mathbb{A}_{y, (m,n)}(x)\, y^m z^n\,,
&
\\
&
\mathbb{A}_{z}(x, y, z) = \frac{\mathbb{A}_{z, (0,-1)}}{z} + \sum_{m, n = 0}^{+\infty} \mathbb{A}_{z, (m,n)}(x)\, y^m z^n\,,
&
\end{aligned}
\right.
\end{equation}
where $\mathbb{A}_{y, (-1,0)}$ and  $\mathbb{A}_{z, (0,-1)}$ are constant matrices. In the lowest-order approximation,
\begin{equation}
\begin{aligned}
d\mathbb{A}(x, y, z)
& =
\mathbb{A}_{x,(0,0)}(x)\, dx
+
\frac{\mathbb{A}_{y, (-1,0)}}{y}\, dy
+
\frac{\mathbb{A}_{z, (0,-1)}}{z}\, dz
\\
& =
\mathbb{A}_{y, (-1,0)}\, d\log y + \mathbb{A}_{z, (0,-1)}\, d\log z
+
\sum_{i = 1, 4, \ldots, 7}\, \widetilde{\mathbb{C}}_{i}\, d\log \widetilde{\omega}_{i}(x)\,.
\end{aligned}
\end{equation}
The symbol alphabet of this approximated canonical differential system comprises $7$ letters,
\begin{align}
\widetilde{\omega}_{1} & = x\,,  &
\widetilde{\omega}_{4} & = x + 1\,,  &
\widetilde{\omega}_{7} & = (x - r_1)/(x + r_1)\,,
\nonumber \\
\widetilde{\omega}_{2} & = y\,,  &
\widetilde{\omega}_{5} & = x - 1\,,
\\
\widetilde{\omega}_{3} & = z\,,  &
\widetilde{\omega}_{6} & = x + 4\,,  &
\nonumber
\end{align}
where the sole square root $r_1$ can be rationalized by the transformation $x \longmapsto x_2$ as defined in eq. \eqref{eq:XYZchange}. Thus, the solution of this approximated differential system, commonly referred to as the lowest-order solution, can be derived using the standard approach for solving canonical differential systems with rational coefficient matrices, along with the following boundary conditions:
\begin{itemize}
\item $g_{1, 4, 9, 10}$ are known from ref. \cite{DiVita:2017xlr}.
\item $g_{2, 5, 11, \ldots, 17, 27, 28}$ are matched with $g_{1, 2, 6, 7, 8, 9, 11, 12, 13, 22, 23}$ of $\mathcal{T}_{1\digamma}$.
\item The remaining $30$ undetermined integration constants are fixed by applying regularity conditions at selected spurious singularities.
      \begin{itemize}
      \item regularity at $x = 0$: $g_{3, 8, 18, \ldots, 26, 29, 30, 32, 35, 36, 37, 42, 43, 44, 45}$
      \item regularity at $y = 0$: $g_{29, 37, 41, 42, 44, 45}$
      \item regularity at $z = 0$: $g_{6, 32}$
      \item regularity at $x = -\, 4$: $g_{39}$
      \end{itemize}
\end{itemize}
The leading behavior of $\bm{g}$ near $y = z = 0$ is well-represented by its lowest-order approximation, $\bm{g}_{\scriptscriptstyle{\text{LO}}}$. The regularity of $\bm{g}$ at $y = 0$ and $z = 0$, as inferred from the method of expansion by regions, implies that $\bm{g}$ is finite at $(x, y, z) = (x, 0, 0)$. Consequently,
\begin{equation}
\bm{g}(x, 0, 0; \epsilon) = \bm{g}_{\scriptscriptstyle{\text{LO}}}(x, 0, 0; \epsilon)\,,
\end{equation}
which can be expressed in terms of GPLs with constant weight vectors and the variable $x_2$.

\par
Based on the above discussion, it is recommended to set $(x, y, z) = (x, 0, 0)$ as the initial point $\bm{x}_0$ for the integration path of $\bm{g}$, at which the boundary value $\bm{g}(\bm{x}_{0}; \epsilon)$ can be obtained using the lowest-order approximation. The integration path is chosen to be a straight line in the $(x_2, y_2, z_2)$ parameter space, connecting the initial point $\bm{x}_0$ to the point of interest $\bm{x}$. Since $y_2 = 0$ and $z_2 = -1$ at the initial point $\bm{x}_0$, the integration path can be parameterized as follows:
\begin{equation}
\gamma:
\quad
\left\lbrace~
\begin{aligned}
&
x_2(\kappa) = x_2
\\
&
y_2(\kappa) = \kappa\, y_2
\\
&
z_2(\kappa) = \kappa\, (z_2 + 1) - 1
\end{aligned}
\right.
\qquad
0 \leqslant \kappa \leqslant 1\,.
\end{equation}
We expand the path-ordered integral as a Taylor series in $\epsilon$,
\begin{equation}
\bm{g}(\bm{x}; \epsilon) = \sum_{n=0}^{+\infty}\, \epsilon^n\, \bm{g}^{(n)}(\bm{x})\,,
\end{equation}
where $\bm{g}^{(n)}(\bm{x})~ (n \in \mathbb{N})$ are determined iteratively by
\begin{equation}
\label{eq:ChenInt1}
\bm{g}^{(n)}(\bm{x})
=
\int_{\gamma} d\mathbb{A}\, \bm{g}^{(n-1)} + \bm{g}^{(n)}(\bm{x}_0)\,.
\end{equation}
By applying integration by parts \cite{Bonciani:2016qxi}, $\bm{g}^{(n)}$ can be written as a one-fold integral over $\bm{g}^{(n-2)}$,
\begin{equation}
\label{eq:ChenInt2}
\bm{g}^{(n)}(\bm{x})
=
\int_{\gamma} \Big[\, \mathbb{A}(\bm{x}) - \mathbb{A} \,\Big]\, d\mathbb{A}\, \bm{g}^{(n-2)}
+
\Big[\, \mathbb{A}(\bm{x}) - \mathbb{A}(\bm{x}_0) \,\Big]\, \bm{g}^{(n-1)}(\bm{x}_0)
+
\bm{g}^{(n)}(\bm{x}_0)\,.
\end{equation}
Up to $\mathcal{O}(\epsilon^2)$, all $45$ MIs can be expressed in terms of GPLs. By applying eqs. \eqref{eq:ChenInt1} and \eqref{eq:ChenInt2}, these MIs at the third and fourth orders in $\epsilon$ can be represented as one-fold integrals over GPLs. Notably, these one-fold integrals can be further simplified to GPLs, except for $g_{36}$ at the third order and $g_{36, 37, 38, 44, 45}$ at the fourth order.

\par
To simplify our analytic expressions, we isolate a closed, self-contained, and rationalizable subsystem from the full canonical differential system. This subsystem comprises $33$ basis integrals, namely $g_{1, \ldots, 28, 32, 33, 34, 35, 39}$. We recalculate these MIs within the subsystem. Compared to the full differential system, the subsystem involves only two square roots: $r_1$ and $r_2$. After applying the two-dimensional transformation $(x, z) \longmapsto (x_2, z_2)$ in eq. \eqref{eq:XYZchange}, the two square roots are rationalized. In the subsequent path-ordered integration for this subsystem, the integration path is chosen to be a straight line in the $(x_2, y, z_2)$ parameter space for convenience. To demonstrate the advantage of introducing a subsystem, we present the first-order expressions of $g_{13}$, derived independently from both the $45$-dimensional full differential system and the $33$-dimensional subsystem, as follows:
\begin{align}
\label{eq:g13}
g_{13\, (\text{full})}^{(1)}
=\,&
- 2\, G(a_{7}; 1)
+ G(a_{8}; 1)
+ G(a_{9}; 1)
- G(a_{10}; 1)
- G(a_{11}; 1)
+ G(a_{12}; 1)
+ G(a_{13}; 1)\,,
&
\nonumber \\
g_{13\, (\text{sub})}^{(1)}
=\,&
\,G(a_{6}; 1)\,,
&
\end{align}
where the weights $a_{6, \ldots, 13}$ are defined in eqs. \eqref{eq:weight-T2-1} and \eqref{eq:weight-T2-2}. It is evident that $g_{13\, (\text{full})}^{(1)}$ contains seven distinct GPLs, while $g_{13\, (\text{sub})}^{(1)}$ involves only one. The overall efficiency of the subsystem is quantified by table \ref{tab:NGPL}, which lists the total numbers of distinct GPLs used in the $33$ MIs $g_{1, \ldots, 28, 32, 33, 34, 35, 39}$ at different orders of $\epsilon$, for both the full differential system and its subsystem. We observe that the higher the order of $\epsilon$, the greater the efficiency of the subsystem. Therefore, for $g_{1, \ldots, 28, 32, 33, 34, 35, 39}$, we adopt the more concise expressions derived from the subsystem. In appendix \ref{appendix:B2}, we present the explicit expressions for the canonical MIs $g_i~ (i = 1, \ldots, 45)$ of the top-branch $\mathcal{T}_{2\digamma}$ up to $\mathcal{O}(\epsilon^2)$. The analytic expressions for the $45$ canonical MIs up to $\mathcal{O}(\epsilon^4)$ are available in the supplementary file ``analytic\_T2.m.''
\begin{table}[htbp]
\centering
\renewcommand{\arraystretch}{1}
\begin{tabular}{
m{1.5cm}<{\centering}
m{3.5cm}<{\centering}
m{3.5cm}<{\centering}
}
\toprule[1.5pt]
Order & $N_{\text{GPL}}$ (full system) & $N_{\text{GPL}}$ (subsystem) \\
\midrule[1pt]
$\epsilon^1$ & $9$ & $6$ \\
$\epsilon^2$ & $104$ & $59$ \\
$\epsilon^3$ & $1702$ & $678$ \\
$\epsilon^4$ & $26440$ & $6511$ \\
\bottomrule[1.5pt]
\end{tabular}
\caption{
\label{tab:NGPL}
Total numbers of distinct GPLs appearing in the analytic expressions of the $33$ MIs $g_{1, \ldots, 28, 32, 33, 34, 35, 39}$, derived independently from both the full differential system and its subsystem, at different orders of $\epsilon$. 
}
\end{table}

\subsection{Top-branch $\mathcal{T}_{3\digamma}$}
\label{subsec:3.3}
\par
\begin{figure}[htbp]
\centering
\includegraphics[width=1\textwidth]{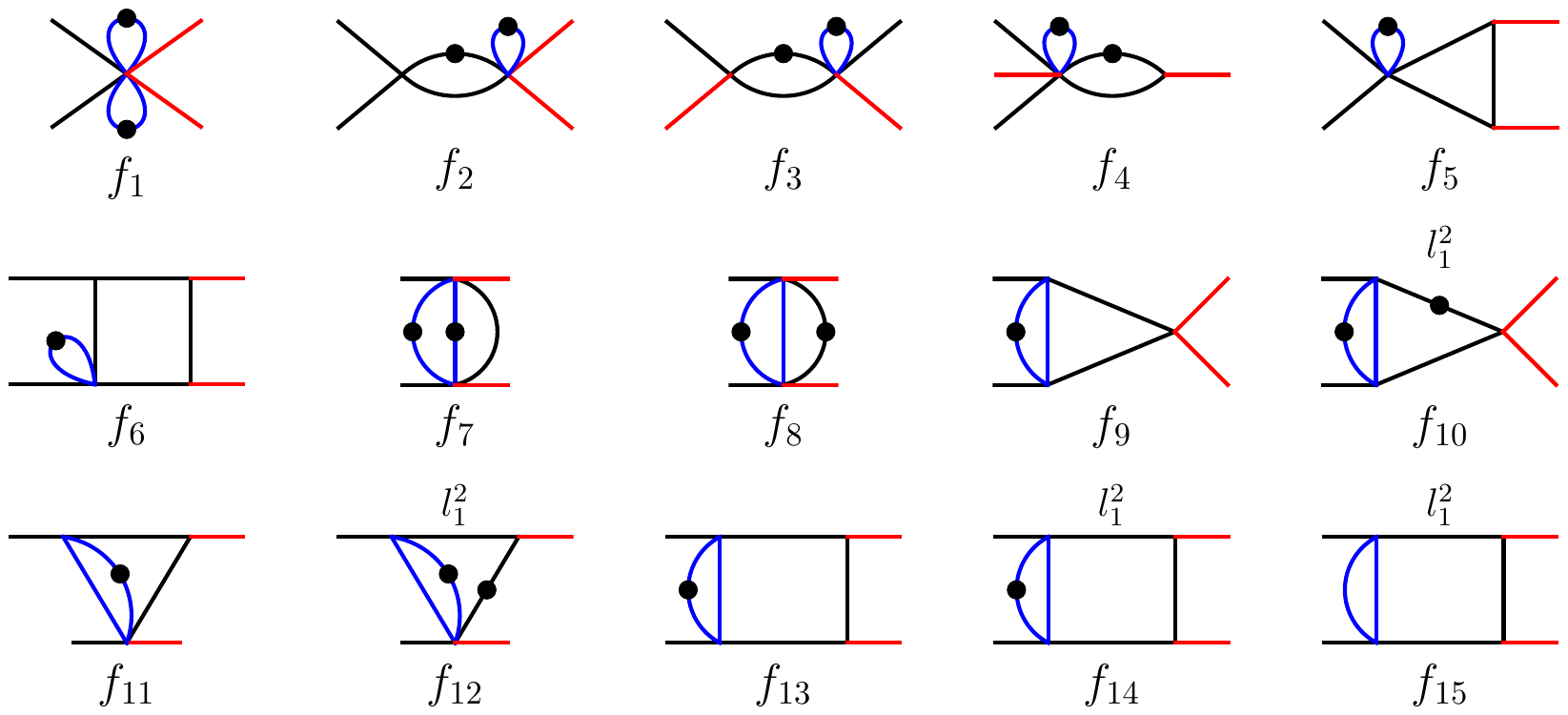}
\caption{
A basis of MIs for the top-branch $\mathcal{T}_{3\digamma}$.
}
\label{fig4}
\end{figure}

The basis of MIs of the top-branch $\mathcal{T}_{3\digamma}$, $\bm{f} = (f_{1}, \ldots, f_{15})^{T}$, is shown in figure \ref{fig4}. The primary challenge in solving $\mathcal{T}_{3\digamma}$ arises from the presence of the elliptic sector $\mathcal{T}_{\text{elliptic}}$,
\begin{equation}
\mathcal{T}_{\text{elliptic}}
=
[\,0,\, 1,\, 1,\, 1,\, 0,\, 0,\, 1,\, 1,\, 0\,]\,.
\end{equation}
Achieving a canonical form for the differential system of $\mathcal{T}_{3\digamma}$ is elusive due to the involvement of elliptic Feynman integrals. To analyze this elliptic sector, we calculate the maximal cut \cite{Kosower:2011ty,Lee:2012te,Primo:2016ebd,Bosma:2017ens,Primo:2017ipr} of the integral $f_{13} = F(0, 1, 1, 2, 0, 0, 1, 1, 0) \in \mathcal{T}_{\text{elliptic}}$ in the loop-by-loop Baikov representation \cite{Baikov:1996iu,Frellesvig:2017aai,Harley:2017qut,Chen:2022lzr}, and then obtain
\begin{equation}
\label{eq:maxcut}
\text{MaxCut} \big(f_{13}\big)
=
\frac{1}{4 \pi^3 \sqrt{x\, (x - 4\, z)}}
\intop_{\mathcal{C}_{\text{MaxCut}}}
\frac{d\xi}{\sqrt{(\xi - \xi_1)\, (\xi - \xi_2)\, (\xi - \xi_3)\, (\xi - \xi_4)}} + \mathcal{O}(\epsilon)
\end{equation}
with
\begin{equation}
\xi_{1,2} = -\, \frac{(x - 2\, z)\, y + 2\, z^2 \pm 2\, z\, \sqrt{x\, y + (y - z)^2}}{x - 4\, z}\,,
\qquad
\xi_3 = 0\,,
\qquad
\xi_4 = 4\,.
\end{equation}
The integrand of the maximal cut in eq. \eqref{eq:maxcut} reveals that the elliptic curve associated with $\mathcal{T}_{\text{elliptic}}$ is characterized by the following quartic polynomial:
\begin{equation}
(\xi, \vartheta): 
\quad
\vartheta^2 = (\xi - \xi_1)\, (\xi - \xi_2)\, (\xi - \xi_3)\, (\xi - \xi_4)\,.
\end{equation}
The modulus $k$ and the complementary modulus $\bar{k}$ of this elliptic curve are defined by
\begin{equation}
k^2 = \frac{U_1}{U_3}\,,
\qquad\qquad
\bar{k}^2 = \frac{U_2}{U_3}
\end{equation}
with
\begin{equation}
U_1 = (\xi_3 - \xi_2)\, (\xi_4 - \xi_1)\,,
\quad
U_2 = (\xi_2 - \xi_1)\, (\xi_4 - \xi_3)\,,
\quad
U_3 = (\xi_3 - \xi_1)\, (\xi_4 - \xi_2)\,.
\end{equation}
Accordingly, the two independent periods of the elliptic curve are
\begin{equation}
\Psi_1= \frac{4\, K(k)}{U_3^{1/2}}\,,
\qquad\qquad
\Psi_2 = \frac{4\, i\, K(\bar{k})}{U_3^{1/2}}\,,
\end{equation}
where $K(x)$ is the complete elliptic integral of the first kind,
\begin{equation}
K(x) = \int_0^1 \frac{dt}{\sqrt{(1 - t^2)\, (1 - x^2\, t^2)}}\,.
\end{equation}

\par
By leveraging the periods of the elliptic curve, we can construct a new basis of MIs that satisfies the linear-form differential equations in eq. \eqref{eq:LDEs}, with $\mathbb{A}_i^{(0)}(\bm{x})$ being strictly lower triangular. This linear basis of MIs is given by
\begin{align}
\label{eq:LBasis}
g_1 & = \epsilon^2\, f_1\,,  &
g_9 & = \epsilon^3\, f_9\, x\,,
\nonumber \\
g_2 & = \epsilon^2\, f_2\, x\,,  &
g_{10} & = \epsilon^2\, f_{10}\, r_1^{\prime} - \epsilon^3\, f_9\, r_1^{\prime}\,,
\nonumber \\
g_3 & = \epsilon^2\, f_3\, y\,,  &
g_{11} & = \epsilon^3\, f_{11}\, (y - z)\,,
\nonumber \\
g_4 & = \epsilon^2\, f_4\, z\,,  &
g_{12} & = \big[\, \epsilon^2\, f_{12}\, z + \epsilon^3\, f_{11}\, (y - z) - \epsilon^2\, f_8\, y\, \big] \, r_4^{\prime} \big/ (y - z)\,,
\nonumber \\
\begin{aligned}
g_5 & \vphantom{= \epsilon^3\, f_5\, r_3^{\prime}}
\\
g_6 & \vphantom{= \epsilon^3\, f_6\,x\, y}
\\
g_7 & \vphantom{= \epsilon^2\, f_7\, y}
\\
g_8 & \vphantom{= \epsilon^2\, f_8\, r_2^{\prime} + 1/2\, f_7\, r_2^{\prime}}
\end{aligned}
&
\begin{aligned}
& = \epsilon^3\, f_5\, r_3^{\prime}\,,
\\
& = \epsilon^3\, f_6\,x\, y\,,
\\
& = \epsilon^2\, f_7\, y\,,
\\
& = \epsilon^2\, f_8\, r_2^{\prime} + 1/2\, f_7\, r_2^{\prime}\,,
\end{aligned}
&
\begin{aligned}
g_{13} & \vphantom{= \epsilon^3\, \frac{\pi\, r_3^{\prime}}{\Psi_1}\, f_{13}}
\\
g_{14} & \vphantom{= \epsilon^3\, f_{14}\, r_3^{\prime}}
\\
g_{15} & \vphantom{= \frac{1}{\epsilon}\, \frac{\Psi_1^2}{2\, \pi\, i\, W_y}\, \frac{\partial g_{13}}{\partial y}}
\end{aligned}
&
\begin{aligned}
& = \epsilon^3\, \frac{\pi\, r_3^{\prime}}{\Psi_1}\, f_{13}\,,
\\
& = \epsilon^3\, f_{14}\, r_3^{\prime}\,,
\\
& = \frac{1}{\epsilon}\, \frac{\Psi_1^2}{2\, \pi\, i\, W_y}\, \frac{\partial g_{13}}{\partial y}\,,
\end{aligned}
\end{align}
where $W_y$ is the Wronskian of $\{\Psi_1, \Psi_2\}$, defined as
\begin{equation}
W_y
=
\Psi_1\, \frac{\partial \Psi_2}{\partial y} - \Psi_2\, \frac{\partial \Psi_1}{\partial y}\,,
\end{equation}
and can be explicitly written as
\begin{equation}
W_y
=
\frac{4\, \pi\, i}{U_3}\, \frac{\partial}{\partial y} \log \frac{U_2}{U_1}
=
\frac{2\, \pi\, i\, (x - 4\, z)\, \big[\, x\, (3\, y + 4) + 2\, (y - z)\, (y - 2\, z) - 16\, z \,\big]}{y\, \big[\, x\, y + (y - z)^2 \,\big]\, \big[\,16\, z\, (y - z + 4) - x\, (y + 4)^2 \,\big]}\,.
\end{equation}
The four distinct square roots $r_{1, 2, 3, 4}^{\prime}$ in eq. \eqref{eq:LBasis} are defined by
\begin{align}
r_1^{\prime\, 2} & = x\, (x - 4)\,,  &
r_2^{\prime\, 2} & = y\, (y + 4)\,,
\nonumber \\
r_3^{\prime\, 2} & = x\, (x - 4\, z)\,,  &
r_4^{\prime\, 2} & = (z - y)\, (y - z + 4)\,.
\end{align}
The linear basis $\bm{g}$, which is the solution of the linear-form differential system \eqref{eq:LDEs}, can be expressed in an iterated form due to the strictly lower-triangular structure of $\mathbb{A}_i^{(0)}(\bm{x})$. Despite the involvement of elliptic Feynman integrals in the top-branch $\mathcal{T}_{3\digamma}$, we can still apply the strategy proposed in subsection \ref{subsec:3.2} to calculate the values of $g_i(\bm{x}; \epsilon)$ at an appropriate initial point. In our calculation, we choose $(x, y, z) = (x, 0, 1)$ as the initial point $\bm{x}_0$ for the following two reasons:
\begin{enumerate}
\item All $15$ MIs of the linear basis $\bm{g}$ are regular at this point.
\item The lowest-order approximation of $\mathcal{T}_{3\digamma}$ near this point is a rationalizable canonical differential system.
\end{enumerate}

\par
To elaborate, the lowest-order approximation of the coefficient matrices $\mathbb{A}_{i}^{(a)}(\bm{x})~ (a = 0, 1~ \text{and}~ i = x, y, z)$ in eq. \eqref{eq:LDEs} for $\mathcal{T}_{3\digamma}$ takes the following form near $(y, z) = (y_0, z_0)$:
\begin{equation}
\label{eq:LAT3}
\mathbb{A}_{x}^{(a)}(\bm{x}) = \mathbb{A}_{x, (0,0)}^{(a)}(x)\,,
\qquad
\mathbb{A}_{y}^{(a)}(\bm{x}) = \frac{\mathbb{A}_{y, (-1,0)}^{(a)}}{y - y_0}\,,
\qquad
\mathbb{A}_{z}^{(a)}(\bm{x}) = \frac{\mathbb{A}_{z, (0,-1)}^{(a)}}{z - z_0}\,.
\end{equation}
Specifically, when $(y_0, z_0) = (0, 1)$, four of the six matrices in eq. \eqref{eq:LAT3} become zero,
\begin{equation}
\mathbb{A}_{x, (0,0)}^{(0)}(x) =
\mathbb{A}_{y, (-1,0)}^{(0)} =
\mathbb{A}_{z, (0,-1)}^{(0)} =
\mathbb{A}_{z, (0,-1)}^{(1)} = 0\,.
\end{equation}
As a result, the approximated differential system reduces to the following canonical form:
\begin{equation}
\label{eq:CanonicalDE3}
d \bm{g}(\bm{x}; \epsilon)
=
\epsilon\,
\Big[\,
\mathbb{A}_{x, (0,0)}^{(1)}(x)\, dx + \frac{1}{y}\, \mathbb{A}_{y, (-1,0)}^{(1)}\, dy
\,\Big]\,
\bm{g}(\bm{x}; \epsilon)\,.
\end{equation}
Moreover, the six irrational functions involved in the coefficient matrices simplify to
\begin{align}
\label{eq:IR6}
\begin{aligned}
& r_1^{\prime} = r_3^{\prime} = \sqrt{x\, (x - 4)}\,,
\\
& r_2^{\prime} = 0\,,
\\
& r_4^{\prime} = \sqrt{3}\,,
\end{aligned}
\qquad\qquad
\begin{aligned}
& \Psi_1 = \frac{\pi}{2}\, \sqrt{x - 4}\,,
\\
& \partial_y \Psi_1 = \frac{\pi}{8}\, (2 - x)\, \sqrt{x - 4}\,.
\end{aligned}
\end{align}
We observe that only two square roots, $\sqrt{x-4}$ and $\sqrt{x\, (x - 4)}$, appear in eq. \eqref{eq:IR6}, both of which can be simultaneously rationalized by applying the change of variables
\begin{equation}
x = \Big( x_3 + \frac{1}{x_3} \Big)^2\,.
\end{equation}
Therefore, the linear-form differential system of $\mathcal{T}_{3\digamma}$ can be approximated as a rationalizable canonical differential system in the vicinity of $(y, z) = (0, 1)$. The solution to eq. \eqref{eq:CanonicalDE3} can be expressed simply in terms of GPLs and is specified by the following boundary conditions:
\begin{itemize}
\item $g_{1,\ldots, 8}$ are taken from ref. \cite{DiVita:2017xlr}.
\item $g_{11} = -\, g_{9}$ and $g_{12} = i\, g_{10}$ at $(x, y, z) = (1, 0 , 1)$.
\item The remaining $5$ integration constants are determined by the regularity conditions as follows.
      \begin{itemize}
      \item $g_{9}$ is regular at $x = 0$.
      \item $g_{13}$ is regular at $x = 0$, $x = 3$, and $y = 0$.
      \item $g_{14}$ is regular at $x = 4$.
      \end{itemize}
\end{itemize}

\par
Returning to the original linear-form differential system of $\mathcal{T}_{3\digamma}$, the values of the MIs at the initial point $\bm{x}_0 = (x, 0, 1)$ are identical to those obtained from the approximated differential system, i.e.,
\begin{equation}
\bm{g}(\bm{x}_0; \epsilon) = \bm{g}_{\scriptscriptstyle{\text{LO}}}(\bm{x}_0; \epsilon)\,.
\end{equation}
The integration path is chosen as a straight line connecting $\bm{x}_0$ to $\bm{x}$ in the $(x_3, y, z)$ parameter space,
\begin{equation}
\gamma:
\quad
\left\lbrace
\begin{aligned}
x_3(\kappa) &= x_3
\\
y(\kappa) &= \kappa\, y
\\
z(\kappa) &= \kappa\, (z - 1) + 1
\end{aligned}
\right.
\qquad
0 \leqslant \kappa \leqslant 1\,.
\end{equation}
Eventually, the MIs $g_i~ (i = 1, \ldots, 15)$ of $\mathcal{T}_{3\digamma}$ are expressed in terms of iterated integrals. The analytic expressions of all the $15$ MIs are given in the supplementary file ``analytic\_T3.m.'' In appendix \ref{appendix:B3}, we only showcase all three MIs belonging to the elliptic sector.

\section{Numerical checks}
\label{sec:4}
\par
The analytic expressions for the canonical bases of $\mathcal{T}_{1\digamma}$ and $\mathcal{T}_{2\digamma}$, as well as the linear basis of $\mathcal{T}_{3\digamma}$, are provided as supplementary material in an electronic format appended to this paper. As previously discussed, the canonical MIs of $\mathcal{T}_{1\digamma}$ and $\mathcal{T}_{2\digamma}$, up to $\mathcal{O}(\epsilon^4)$, can be expressed either as GPLs or as one-fold integrals over GPLs. For the symbolic computation and numerical evaluation of GPLs, we utilize the Mathematica package \texttt{PolyLogTools} \cite{Maitre:2005uu,Maitre:2007kp,Duhr:2019tlz} and the C++ library \texttt{GiNaC} \cite{Bauer:2000cp,Vollinga:2004sn}. To validate the correctness of our analytic solutions to the differential systems of $\mathcal{T}_{1\digamma}$ and $\mathcal{T}_{2\digamma}$, we perform cross-verification against the numerical results obtained using \texttt{AMFlow} \cite{Liu:2017jxz,Liu:2022chg} in the Euclidean region $\{\, (x, y, z) \,|\, x > 0,\, y > 0,\, z > 0 \,\}$. In table \ref{tab:checkT12}, we present a comparison between the numerical results obtained from our analytic expressions and those from \texttt{AMFlow} for selected MIs of $\mathcal{T}_{1\digamma}$ and $\mathcal{T}_{2\digamma}$ at $(x, y, z) = (17/2, 1/3, 5/3)$. Notably, $g_{25}$ and $g_{42}$ in table \ref{tab:checkT12} are representative MIs of the top-sectors of the first two family branches, respectively,
\begin{equation}
g_{25} \in \mathcal{T}_1 \subset \mathcal{T}_{1\digamma}\,,
\qquad\qquad
g_{42} \in \mathcal{T}_2 \subset \mathcal{T}_{2\digamma}\,.
\end{equation}
Both can be expressed in terms of GPLs. In contrast, $g_{36} \in \mathcal{T}_{2\digamma}$ is an exemplary MI that can be formulated as a one-fold integral over GPLs. The numerical results obtained from our analytic formulations exhibit exceptional agreement with those computed using the auxiliary mass flow method, demonstrating an extremely high level of accuracy, as evidenced by
\begin{equation}
\left|\,
\big( I_{\text{analytic}} - I_{\text{AMFlow}} \big) \big/ I_{\text{AMFlow}}
\,\right|
< 10^{-38}\,.
\end{equation}
\begin{table}[htbp]
\centering
\renewcommand{\arraystretch}{1.4}
\begin{tabular}{
p{1.8cm}<{\centering}
p{1.8cm}<{\centering}
l
}
\toprule[1.5pt]
\multirow{1}{*}{Branch}
&
\multirow{1}{*}{$\textbf{G}$}
&
\qquad\qquad\qquad
\multirow{1}{*}{Results ( Analytic \big/ \texttt{AMFlow} )}
\\
\midrule[1pt]
\multirow{5.5}{*}{$\mathcal{T}_{1\digamma}$}
&
\multirow{5.3}{*}{$g_{25}$}
&
$
\begin{aligned}
-\,&
0.25
\\
+\,&
0.3768859011881900759989191267492984156809\, \epsilon
\\
-\,&
1.2620443490084127732359725216242483420669\, \epsilon^2
\\
-\,&
0.7517989880793716026311279513267280441939\, \epsilon^3
\\
+\,&
2.5243257885032819715982812514126356527038\, \epsilon^4
\end{aligned}
$
\\
\cmidrule{3-3}
&
&
$
\begin{aligned}
-\,&
0.25
\\
+\,&
0.3768859011881900759989191267492984156809\, \epsilon
\\
-\,&
1.2620443490084127732359725216242483420669\, \epsilon^2
\\
-\,&
0.7517989880793716026311279513267280441939\, \epsilon^3
\\
+\,&
2.5243257885032819715982812514126356527038\, \epsilon^4
\end{aligned}
$
\\
\midrule[1pt]
\multirow{9.1}{*}{$\mathcal{T}_{2\digamma}$}
&
\multirow{3.6}{*}{$g_{42}$}
&
$
\begin{aligned}
-\,&
2.1153713978990082749586947832894060646047\, \epsilon^2
\\
-\,&
0.8166254210332590669598372077758470682941\, \epsilon^3
\\
+\,&
1.8108236510538266812815390892429259757628\, \epsilon^4
\end{aligned}
$
\\
\cmidrule{3-3}
&
&
$
\begin{aligned}
-\,&
2.1153713978990082749586947832894060646047\, \epsilon^2
\\
-\,&
0.8166254210332590669598372077758470682941\, \epsilon^3
\\
+\,&
1.8108236510538266812815390892429259757628\, \epsilon^4
\end{aligned}
$
\\
\cmidrule{2-3}
&
\multirow{2.9}{*}{$g_{36}$}
&
$
\begin{aligned}
&
0.9282065930624508142424433019313325832520\, \epsilon^3
\\
+\,&
1.2458618319714537281221371673463469359882\, \epsilon^4
\end{aligned}
$
\\
\cmidrule{3-3}
&
&
$
\begin{aligned}
&
0.9282065930624508142424433019313325832520\, \epsilon^3
\\
+\,&
1.2458618319714537281221371673463469359906\, \epsilon^4
\end{aligned}
$
\\
\bottomrule[1.5pt]
\end{tabular}
\caption{
\label{tab:checkT12}
Comparison between the numerical results obtained from our analytic expressions and \texttt{AMFlow} for $g_{25} \in \mathcal{T}_{1\digamma}$ and $g_{42, 36} \in \mathcal{T}_{2\digamma}$ at $(x, y, z) = (17/2, 1/3, 5/3)$.
}
\end{table}

\par
The third family branch features an elliptic sector, resulting in the MIs of $\mathcal{T}_{3\digamma}$ being represented by iterated integrals. Due to the intricate nature of the integration kernels, computing these MIs requires substantial computational resources and effort. Consequently, there is a pressing need to devise a more efficient approach for computation. To verify the correctness of our iterated solution, we compare its numerical outcomes with those obtained using \texttt{AMFlow} for $g_{13,14,15} \in \mathcal{T}_{\text{elliptic}} \subset \mathcal{T}_{3\digamma}$, as detailed in table \ref{tab:checkT3}. The comparison reveals a consistent match, with discrepancies within an accuracy of $10^{-15}$.
\begin{table}[htbp]
\centering
\renewcommand{\arraystretch}{1.4}
\begin{tabular}{
p{1.8cm}<{\centering}
p{1.8cm}<{\centering}
l
p{1.8cm}<{\centering}
}
\toprule[1.5pt]
\multirow{1}{*}{Branch}
&
\multirow{1}{*}{$\textbf{G}$}
&
\quad
\multirow{1}{*}{Results ( Analytic \big/ \texttt{AMFlow} )}
\\
\midrule[1pt]
\multirow{6.5}{*}{$\mathcal{T}_{3\digamma}$}
&
\multirow{1}{*}{$g_{13}$}
&
$
\begin{aligned}
~~~&
3.2548275606982903057\, \epsilon^3
\\
&
3.2548275606982903051\, \epsilon^3
\end{aligned}
$
&
$
\begin{aligned}
&
\scriptstyle{\text{(Analytic)}}
\\
&
\scriptstyle{\text{(AMFlow)}}
\end{aligned}
$
\\
\cmidrule{2-4}
&
\multirow{1}{*}{$g_{14}$}
&
$
\begin{aligned}
-\,&
2.33756055601827687\, \epsilon^3
\\
-\,&
2.33756055601827691\, \epsilon^3
\end{aligned}
$
&
$
\begin{aligned}
&
\scriptstyle{\text{(Analytic)}}
\\
&
\scriptstyle{\text{(AMFlow)}}
\end{aligned}
$
\\
\cmidrule{2-4}
&
\multirow{1}{*}{$g_{15}$}
&
$
\begin{aligned}
-\,&
0.094132461909778403737363\, \epsilon^2
\\
-\,&
0.0715655314006887\, \epsilon^3
\\
-\,&
0.094132461909778403737361\, \epsilon^2
\\
-\,&
0.0715655314006885\, \epsilon^3
\end{aligned}
$
&
$
\begin{aligned}
&
\scriptstyle{\text{(Analytic)}}
\\
&
\\
&
\scriptstyle{\text{(AMFlow)}}
\end{aligned}
$
\\
\bottomrule[1.5pt]
\end{tabular}
\caption{
\label{tab:checkT3}Comparison between the numerical results obtained from our analytic expressions and \texttt{AMFlow} for $g_{13,14,15} \in \mathcal{T}_{3\digamma}$ at $(x, y, z) = (17/2, 1/3, 5/3)$.}
\end{table}

\section{Summary}
\label{sec:5}
\par
Vector-boson pair production in high-energy collisions plays a pivotal role in studying the electroweak symmetry breaking mechanism. In this paper, we present an analytic calculation of the planar two-loop MIs for the massive NNLO QCD corrections to $W$-boson pair production in quark-antiquark annihilation. These MIs, which form the bases of three distinct family branches, satisfy three different systems of differential equations. For the first family branch, the differential equations can be cast into a canonical form with rational coefficient matrices by choosing an appropriate basis of MIs and employing suitable kinematic variables, thereby enabling the solution to be elegantly expressed in terms of GPLs. For the second family branch, the differential equations can also be transformed into canonical form; however, the four square roots involved cannot be simultaneously rationalized. Fortunately, the MIs of this family branch can be expressed either as GPLs or as one-fold integrals over GPLs, up to $\mathcal{O}(\epsilon^4)$. Compared to the first two family branches, the third one is more complicated due to the involvement of elliptic Feynman integrals. By leveraging the periods of the associated elliptic curve, we construct a set of MIs that satisfies linear-form differential equations. This linear basis of MIs can be expressed in an iterated form, owing to the strictly lower-triangular structure of the coefficient matrices at $\epsilon = 0$. The correctness of our analytic expressions for the MIs of all three family branches has been validated against the numerical results from the \texttt{AMFlow} package. Our analytic results can aid in developing a flexible and efficient Monte Carlo program for precise analyses of vector-boson pair production at hadron colliders.

\acknowledgments
\par
This work is supported by the National Natural Science Foundation of China (Grant No. 12061141005) and the CAS Center for Excellence in Particle Physics (CCEPP).

\appendix
\section{Coefficients in the canonical bases of $\mathcal{T}_{1\digamma}$ and $\mathcal{T}_{2\digamma}$}
\label{appendix:A}
\par
The coefficients $\alpha_i$ in eq. \eqref{eq:gT1} are given by
\begin{align}
\alpha_{5} & = -\, 3/2\, \epsilon^2\, x\, (y - z)\,,  &
\alpha_{6} & = 3\, \epsilon^2\, (y - z)\,,
\nonumber \\
\alpha_{7} & = -\, 3/2\, \epsilon^2\, (y - z)\, (z + 1)\,,  &
\alpha_{8} & = \epsilon^2\, (y - z)\, (4\, y + 1)\,,
\nonumber \\
\alpha_{9} & = 1/2\, \epsilon^2\, (y + 1)\, (y - z)\,,  &
\alpha_{14} & = -\, 6\, \epsilon^3\, (y - z)\, \big[\, x - 2\, (z + 1) \,\big]\,,
\nonumber \\
\alpha_{15} & = -\, 4\, \epsilon^2\, (y - z)\, \big[\, x - (z + 1)^2 \,\big]\,,  &
\alpha_{16} & = 6\, \epsilon^3\, (y - z)^2\,,
\\
\alpha_{17} & = -\, 6\, \epsilon^4\, (y - z)\, (x + y - z)\,,  &
\alpha_{18} & = -\, 4\, \epsilon^3\, (y - z)\, \big[\, x + (y - z)\, (z + 1) \,\big]\,,
\nonumber \\
\alpha_{21} & = 2\, \epsilon^3\, x\, (y + 1)\, (z + 1)\,,  &
\alpha_{25} & = -\, \epsilon^4 \, x^2\, (y + 1)\,,
\nonumber \\
\alpha_{27} & = \epsilon^4\, x^2\, (y + 1)\,.  &
\nonumber
\end{align}
The coefficients $\beta_i$ and $\gamma_i$ in eq. \eqref{eq:gT2} are given by
\begin{align}
\beta_{9} & = -\, 3/2\, \epsilon^2\, x\, (y - z)\,,  &
\beta_{11} & = \epsilon^2\, (y - z)\, (2\, z - 1)\,,
\nonumber \\
\beta_{12} & = -\, 1/2\, \epsilon^2\, (y - z)\, (z + 1)\,,  &
\beta_{13} & = -\, \epsilon^2\, (y - z)\, (2\, y - 1)\,,
\nonumber \\
\beta_{14} & = 1/2\, \epsilon^2\, (y + 1)\, (y - z)\,,  &
\beta_{18} & = -\, 3\, \epsilon^3\, (x + 2)\, (y - z)\,,
\nonumber \\
\beta_{19} & = -\, 2\, \epsilon^2\, (x + 2)\, (y - z)\,,  &
\beta_{20} & = -\, 2\, \epsilon^2\, (y - z)\, \big[\, x\, (z + 1) + 1 \,\big]\,,
\\
\beta_{21} & = -\, 3\, \epsilon^3\, (y - z)\, (y - z - 2)\,,  &
\beta_{22} & = -\, 2\, \epsilon^2\, (y - z)\, (y - z - 2)\,,
\nonumber \\
\beta_{23} & = -\, 2\, \epsilon^2\, (y - z)\, (y - z - 1)\,,  &
\beta_{30} & = 1/2\, \epsilon^3 \, x\, (z + 1)\, \big[\, x\, (y + 1) + 2\, (y - z) \big]\,,
\nonumber \\
\beta_{31} & = \rlap{$\displaystyle
                        -\, \epsilon^3\, (y - z)\, (z + 1) \big[\, x\, (z + 1) + (y - z) \,\big]\,,
                        $}  &
\nonumber
\end{align}
and
\begin{align}
\gamma_{7} & = -\, 1/2\, \epsilon^3\, x\,,  &
\gamma_{8} & = -\, 1/2\, \epsilon^3\, x\, \big[\, x + 2\, (y - z) \,\big]\,,
\nonumber \\
\gamma_{18} & = -\, \epsilon^3\, x\,,  &
\gamma_{19} & = -\, \epsilon^2\, x\,,
\nonumber \\
\gamma_{29} & = 2\, \epsilon^4\, (y - z)\,,  &
\gamma_{30} & = -\, 1/2\, \epsilon^3\, x\, (y - z)\,,
\nonumber \\
\gamma_{32} & = \epsilon^4\, x\,,  &
\gamma_{35} & = -\, \epsilon^4\, x\,,
\\
\gamma_{36} & = \epsilon^3\, x\, (y - z + 1)\,,  &
\gamma_{37} & = \epsilon^2\, x\, (y - z)\,,
\nonumber \\
\gamma_{38} & = \epsilon^3\, x\,,  &
\gamma_{39} & = 1/2\, \epsilon^4 \, x\, (x - 2\, z + 2)\,,
\nonumber \\
\gamma_{42} & = -\, 1/2\, \epsilon^4\, x^2\, y\,,  &
\gamma_{43} & = -\, 1/2\, \epsilon^4\, x\, (x - 2\, z)\,,
\nonumber \\
\gamma_{44} & = -\, 1/2\, \epsilon^4\, x^2\,,  &
\gamma_{45} & = \epsilon^4\, x\,.
\nonumber
\end{align}

\section{Explicit expressions for MIs}
\label{appendix:B}
\subsection{Canonical basis of $\mathcal{T}_{1\digamma}$}
\label{appendix:B1}
\par
The canonical MIs $g_{i}~ (i = 1, \ldots, 27)$ of the top-branch $\mathcal{T}_{1\digamma}$ can be expressed in terms of GPLs. The explicit expressions up to $\mathcal{O}(\epsilon^2)$ are listed as follows:
\begin{align}
g_{1} = {} &
                    1
                    - \epsilon\, \big[\,
                    G(a_{1}; x_{1})
                    + 2\, G(a_{4}; z)
                    \,\big]
\nonumber \\
&
                    + \epsilon^{2}\, \big[\,
                    2\, G(a_{1}; x_{1})\, G(a_{4}; z)
                    + G(a_{1}, a_{1}; x_{1})
                    + 4\, G(a_{4}, a_{4}; z)
                    - \pi^{2}/6
                    \,\big]
\nonumber \\
g_{2} = {} &
                    - \epsilon\,  G(a_{3}; z)
\nonumber \\
&
                    +\epsilon^{2}\, \big[\,
                    G(a_{1}; x_{1})\, G(a_{3}; z)
                    - G(a_{1}, a_{3}; z)
                    + 2\, G(a_{3}, a_{3}; z)
                    + 2\, G(a_{3}, a_{4}; z)
                    + 2\, G(a_{4}, a_{3}; z)
                    \,\big]
\nonumber \\
g_{3} = {} &
                    1
                    - 2\, \epsilon\, \big[\,
                    G(a_{1}; x_{1})
                    + 2\, G(a_{4}; z)
                    \,\big]
\nonumber \\
&
                    + 4\, \epsilon^{2}\, \big[\,
                    2\, G(a_{1}; x_{1})\, G(a_{4}; z)
                    + G(a_{1}, a_{1}; x_{1})
                    + 4\, G(a_{4}, a_{4}; z)
                    - \pi^{2}/12
                    \,\big]
\nonumber \\
g_{4} = {} &
                    \epsilon^{2}\, \big[\,
                    G(a_{1}; x_{1})\, G(a_{5}; z)
                    + G(a_{1}; x_{1})\, G(a_{6}; z)
                    - G(a_{2}, a_{1}; x_{1})
                    + 2\, G(a_{1}, a_{3}; z)
\nonumber \\
&
                    - 2\, G(a_{5}, a_{3}; z)
                    + 2\, G(a_{5}, a_{4}; z)
                    - 2\, G(a_{6}, a_{3}; z)
                    + 2\, G(a_{6}, a_{4}; z)                    
                     \, \big]
\nonumber \\
g_{5} = {} &
                    - g_{3}
\nonumber \\
g_{6} = {} & \epsilon\, G(a_{3}; z)
                    + \epsilon^{2}\, \big[\,
                    G(a_{1}, a_{3}; z)
                    - 4\, G(a_{3}, a_{3}; z)
                    \, \big]
\nonumber \\
g_{7} = {} &
                    - 1
                    + 2\, \epsilon\, G(a_{3}; z)
                    + 4\, \epsilon^{2}\, \big[\,
                    G(a_{1}, a_{3}; z)
                    - 2\, G(a_{3}, a_{3};z)
                    - \pi^{2}/12
                    \,\big]
\nonumber \\
g_{8} = {} &
                    \epsilon\, G(a_{3}; y)
                    + \epsilon^{2}\, \big[\,
                    G(a_{1}, a_{3}; y)
                    - 4\, G(a_{3}, a_{3}; y)
                    \,\big]
\nonumber \\
g_{9} = {} &
                    - 1
                    + 2\, \epsilon\, G(a_{3}; y)
                    + 4\, \epsilon^{2}\, \big[\,
                    G(a_{1}, a_{3}; y)
                    - 2\, G(a_{3}, a_{3}; y)
                    - \pi^{2}/12
                    \,\big]
\nonumber \\
g_{10} = {} &
                    g_{3}/4 + \pi^2/12\, \epsilon^2
\nonumber \\
g_{11} = {} & 0
\nonumber \\
g_{12} = {} &
                    - g_{4}/2
\nonumber \\
g_{13} = {} &
                    \epsilon\, G(a_{3}; z)
\nonumber \\
&
                    + \epsilon^{2}/2\,\big[
                    - G(a_{1}; x_{1})\, G(a_{5}; z)
                    - G(a_{1}; x_{1})\, G(a_{6}; z)                    
                    - G(a_{2}, a_{1}; x_{1})
                    + 2\, G(a_{1}, a_{3}; z)
\nonumber \\
&
                    - 8\, G(a_{3}, a_{3}; z)
                    - 4\, G(a_{4}, a_{3}; z)
                    + 2\, G(a_{5}, a_{3}; z)
                    - 2\, G(a_{5}, a_{4}; z)
                    + 2\, G(a_{6}, a_{3}; z)
\nonumber \\
&
                    - 2\, G(a_{6}, a_{4}; z)
                    \,\big]
\nonumber \\
g_{14} = {} &
                    - g_{4}
\nonumber \\
g_{15} = {} &
                    \epsilon\, \big[
                    - G(a_{1}; x_{1})
                    + 2\, G(a_{3}; z)
                    - 2\, G(a_{4}; z)
                    \,\big]
                    + \epsilon^{2}/2\, \big[
\nonumber \\
&
                    - 2\, G(a_{1}; x_{1})\, G(a_{3}; z)
                    + 8\, G(a_{1}; x_{1}) \, G(a_{4}; z)
                    + G(a_{1}; x_{1})\, G(a_{5}; z)
                    + G(a_{1}; x_{1})\, G(a_{6}; z)
\nonumber \\
&
                    + 4\, G(a_{1}, a_{1}; x_{1})
                    + G(a_{2}, a_{1}; x_{1})
                    + 6\, G(a_{1}, a_{3}; z)
                    - 12\, G(a_{3}, a_{3}; z)
                    - 4\, G(a_{3}, a_{4}; z)
\nonumber \\
&
                    + 16\, G(a_{4}, a_{4}; z)
                    - 2\, G(a_{5}, a_{3}; z)
                    + 2\, G(a_{5}, a_{4}; z)
                    - 2\, G(a_{6}, a_{3}; z)
                    + 2\, G(a_{6}, a_{4}; z)
\nonumber \\
&
                    - 2\, \pi^{2}/3
                    \,\big]
\nonumber \\
g_{16} = {} &
                    \epsilon/2\, \big[\,
                    G(a_{3}; z)
                    - G(a_{3}; y)
                    \,\big]
\nonumber \\
&
                    + \epsilon^{2}/2\,\big[
                    - G(a_{7}; y)\, G(a_{3}; z)
                    - 2\, G(a_{1}, a_{3}; y)
                    + 4\, G(a_{3}, a_{3}; y)
                    + G(a_{7}, a_{3}; y)
\nonumber \\
&
                    + G(a_{1}, a_{3}; z)                                       
                    - 3\, G(a_{3}, a_{3}; z)
                    \,\big]
\nonumber \\
g_{17} = {} & 0
\nonumber \\
g_{18} = {} &
                    \epsilon^{2}\, \big[\,
                    G(a_{1}; x_{1})\, G(a_{3}; y)
                    + G(a_{1}; x_{1})\, G(a_{3}; z)
                    - G(a_{1}; x_{1})\, G(a_{5}; z)
                    - G(a_{1}; x_{1})\, G(a_{6}; z)
\nonumber \\
&
                    - 2\, G(a_{3}; y)\, G(a_{3}; z)
                    + 2\, G(a_{3}; y)\, G(a_{4}; z)
                    + G(a_{7}; y)\, G(a_{3}; z)
                    - G(a_{2}, a_{1}; x_{1})
\nonumber \\
&
                    - G(a_{7}, a_{3}; y)
                    + G(a_{1}, a_{3}; z)
                    - G(a_{3}, a_{3}; z)
                    + 2\, G(a_{3}, a_{4}; z)
                    - 2\, G(a_{4}, a_{3}; z)
\nonumber \\
&
                    + 2\, G(a_{5}, a_{3}; z)
                    - 2\, G(a_{5}, a_{4}; z)
                    + 2\, G(a_{6}, a_{3}; z) 
                    - 2\, G(a_{6}, a_{4}; z)
                    \, \big]
\nonumber \\
g_{19} = {} & 0
\nonumber \\
g_{20} = {} &
                    g_{4}
\nonumber \\
g_{21} = {} &
                    - 1/4
                    + \epsilon/2\, \big[\,
                    G(a_{1}; x_{1})
                    + 2\, G(a_{3}; y)
                    - 2\, G(a_{3}; z)
                    + 2\, G(a_{4}; z)
                    \,\big]
                    + \epsilon^{2}\, \big[
\nonumber \\
&
                    - 2\, G(a_{1}; x_{1})\, G(a_{3}; y)
                    - 2\, G(a_{1}; x_{1})\, G(a_{4}; z)
                    + G(a_{1}; x_{1})\, G(a_{5}; z)
                    + G(a_{1}; x_{1})\, G(a_{6}; z)
\nonumber \\
&
                    + 4\, G(a_{3}; y)\, G(a_{3}; z)
                    - 4\, G(a_{3}; y)\, G(a_{4}; z)
                    - 3\, G(a_{7}; y)\, G(a_{3}; z)
                    - G(a_{1}, a_{1}; x_{1})
 \nonumber \\
&
                    + G(a_{2}, a_{1}; x_{1})
                    - 4\, G(a_{3}, a_{3}; y)
                    + 3\, G(a_{7}, a_{3}; y)
                    - 3\, G(a_{1}, a_{3}; z)
                    + 3\, G(a_{3}, a_{3}; z)
\nonumber \\
&
                    + 4\, G(a_{4}, a_{3}; z)
                    - 4\, G(a_{4}, a_{4}; z)
                    - 2\, G(a_{5}, a_{3}; z)
                    + 2\, G(a_{5}, a_{4}; z)
                    - 2\, G(a_{6}, a_{3}; z)
\nonumber \\
&
                    + 2\, G(a_{6}, a_{4}; z)
                     \,\big]
\nonumber \\
g_{22} = {} &
                    \epsilon\, G(a_{3}; y)
\nonumber \\
&
                    + \epsilon^{2}\, \big[
                    - G(a_{1}; x_{1})\, G(a_{3}; y)
                    + 2\, G(a_{3}; y)\, G(a_{3}; z)
                    - 2\, G(a_{3}; y)\, G(a_{4}; z)
\nonumber \\
&
                    - 2\, G(a_{7}; y)\, G(a_{3}; z)
                    - G(a_{1}, a_{3}; y)
                    - 4\, G(a_{3}, a_{3}; y)
                    + 2\, G(a_{7}, a_{3}; y)
                    \,\big]
\nonumber \\
g_{23} = {} &
                    -1/2
                    + \epsilon/2\, \big[\,
                    G(a_{1}; x_{1})
                    + 2\, G(a_{3}; y)
                    - 2\, G(a_{3}; z)
                    + 2\, G(a_{4}; z)
                    \,\big]
                    +\epsilon^{2}/2\, \big[
\nonumber \\
&
                    - 2\, G(a_{1}; x_{1})\, G(a_{3}; y)
                    - 2\, G(a_{1}; x_{1})\, G(a_{4}; z)
                    + 4\, G(a_{3}; y)\, G(a_{3}; z)
                    - 4\, G(a_{3}; y)\, G(a_{4}; z)
\nonumber \\
&
                    - 4\, G(a_{7}; y)\, G(a_{3}; z)
                    + G(a_{1}; x_{1})\, G(a_{5}; z)
                    + G(a_{1}; x_{1})\, G(a_{6}; z)
                    - G(a_{1}, a_{1}; x_{1})
\nonumber \\
&
                    + G(a_{2}, a_{1}; x_{1})
                    + 4\, G(a_{1}, a_{3}; y)
                    - 8\, G(a_{3}, a_{3}; y)
                    + 4\, G(a_{7}, a_{3}; y)
                    - 8\, G(a_{1}, a_{3}; z)
\nonumber \\
&
                    + 8\, G(a_{3}, a_{3}; z)
                    + 4\, G(a_{4}, a_{3}; z)
                    - 4\, G(a_{4}, a_{4}; z)
                    - 2\, G(a_{5}, a_{3}; z)
                    + 2\, G(a_{5}, a_{4}; z)
\nonumber \\
&
                    - 2\, G(a_{6}, a_{3}; z)
                    + 2\, G(a_{6}, a_{4}; z)
                    + \pi^{2}/6
                    \,\big]
\nonumber \\
g_{24} = {} &
                    \epsilon/2\, \big[\,
                    G(a_{3}; z)
                    - G(a_{3}; y)
                    \,\big]
\nonumber \\
&
                    + \epsilon^{2}\, \big[
                    - G(a_{7}; y)\, G(a_{3}; z)
                    - G(a_{1}, a_{3}; y)
                    + 2\, G(a_{3}, a_{3}; y)
                    + G(a_{7}, a_{3}; y)
                    - G(a_{3}, a_{3}; z)
                    \,\big]
\nonumber \\
g_{25} = {} &
                    -1/4
                    + \epsilon/2\, \big[\,
                    G(a_{1}; x_{1})
                    + 2\, G(a_{3}; y)
                    - 2\, G(a_{3}; z)
                    + 2\, G(a_{4}; z)
                    \,\big]
                    + \epsilon^{2}\, \big[
\nonumber \\
&
                    - 2\, G(a_{1}; x_{1})\, G(a_{3}; y)
                    - 2\, G(a_{1}; x_{1})\, G(a_{4}; z)
                    + G(a_{1}; x_{1})\, G(a_{5}; z)
                    + G(a_{1}; x_{1})\, G(a_{6}; z)
\nonumber \\
&
                    + 4\, G(a_{3}; y)\, G(a_{3}; z)
                    - 4\, G(a_{3}; y)\, G(a_{4}; z)
                    - 2\, G(a_{7}; y)\, G(a_{3}; z)                 
                    - G(a_{1}, a_{1}; x_{1})
\nonumber \\
&
                    + G(a_{2}, a_{1}; x_{1})
                    - 4\, G(a_{3}, a_{3}; y)
                    + 2\, G(a_{7}, a_{3}; y)
                    - 2\, G(a_{1}, a_{3}; z)
                    + 2\, G(a_{3}, a_{3}; z)
\nonumber \\
&
                    + 4\, G(a_{4}, a_{3}; z)
                    - 4\, G(a_{4}, a_{4}; z)
                    - 2\, G(a_{5}, a_{3}; z)
                    + 2\, G(a_{5}, a_{4}; z)
                    - 2\, G(a_{6}, a_{3}; z)
\nonumber \\
&
                    + 2\, G(a_{6}, a_{4}; z)                                        
                    - \pi^{2}/6
                    \,\big]
\nonumber \\
g_{26} = {} &
                    \epsilon^{2}\, \big[
                    - G(a_{1}; x_{1})\, G(a_{5}; z)
                    - G(a_{1}; x_{1})\, G(a_{6}; z)                    
                    + G(a_{2}, a_{1}; x_{1})
                    - 2\, G(a_{1}, a_{3}; z)
\nonumber \\
&
                    + 2\, G(a_{5}, a_{3}; z)
                    - 2\, G(a_{5}, a_{4}; z)
                    + 2\, G(a_{6}, a_{3}; z)
                    - 2\, G(a_{6}, a_{4}; z)                                        
                    \,\big]
\nonumber \\
g_{27} = {} &
                    - 1/4
                    + \epsilon/2\, \big[\,
                    G(a_{1}; x_{1})
                    + 2\, G(a_{3}; y)
                    - 2\, G(a_{3};  z)
                    + 2\, G(a_{4}; z)
                    \,\big]
                    + \epsilon^{2}\, \big[
\nonumber \\
&
                    - 2\, G(a_{1}; x_{1})\, G(a_{3}; y)
                    - 2\, G(a_{1}; x_{1})\, G(a_{4}; z)
                    + G(a_{1}; x_{1})\, G(a_{5}; z)
                    + G(a_{1}; x_{1})\, G(a_{6}; z)
\nonumber \\
&
                    + 4\, G(a_{3}; y)\, G(a_{3}; z)
                    - 4\, G(a_{3}; y)\, G(a_{4}; z)
                    - 4\, G(a_{7}; y)\, G(a_{3}; z)                                                          
                    - G(a_{1}, a_{1}; x_{1})
\nonumber \\
&
                    + G(a_{2}, a_{1}; x_{1})
                    - 4\, G(a_{3}, a_{3}; y)
                    + 4\, G(a_{7}, a_{3}; y)
                    - 4\, G(a_{1}, a_{3}; z)
                    + 4\, G(a_{3}, a_{3}; z)
\nonumber \\
&
                    + 4\, G(a_{4}, a_{3}; z)
                    - 4\, G(a_{4}, a_{4}; z)
                    - 2\, G(a_{5}, a_{3}; z)
                    + 2\, G(a_{5}, a_{4}; z)
                    - 2\, G(a_{6}, a_{3}; z)
\nonumber \\
&
                    + 2\, G(a_{6}, a_{4}; z)
                    + \pi^{2}/6
                    \,\big]
\end{align}
where the weights of the GPLs are
\begin{equation}
a_1 = 0\,,
\qquad
a_{2,3} = \pm\, 1\,,
\qquad
a_4 = -\, x_1\,,
\qquad
a_{5,6} = \pm\, \sqrt{x_1}\,,
\qquad
a_7 = z\,.
\end{equation}

\subsection{Canonical basis of $\mathcal{T}_{2\digamma}$}
\label{appendix:B2}
The explicit expressions for the $45$ canonical MIs of the top-branch $\mathcal{T}_{2\digamma}$ up to $\mathcal{O}(\epsilon^2)$ are listed below:

The weights of the GPLs involved in eq. \eqref{eq:Exp-branch2} are given as follows:
\begin{align}
\label{eq:weight-T2-1}
& a_{1} = 0\,,  &
& a_{2} = -\, 1\,, &
& a_{3} = -\, 2\,,  
\nonumber \\
& a_{6} = -\, 1/y\,,  &
& a_{7} = 1/(z_2 + 1)\,,  &
& a_{8} = (x_2 + 1)/(z_2 + 1)\,,  
\\
& a_{9} = 1/\big[\, (x_2 + 1)\, (z_2 + 1) \,\big]\,,  &
& a_{10} = -\, 1/y_2\,,  &
& a_{11} = (x_2 + 1)/y_2\,,  
\nonumber
\end{align}
and $a_{4, 5}$, $a_{12, 13}$, $a_{14, 15}$, $a_{16, 17}$, $a_{18, 19}$, $a_{20, 21}$ are the roots of the following six quadratic equations:
\begin{align}
\label{eq:weight-T2-2}
& a_{4, 5}:  &
& \kappa^2 - \kappa - 1 = 0  &
\nonumber \\
& a_{12, 13}:  &
& y_2^2\, \kappa^2 - (x_2 + 1) = 0  &
\nonumber \\
& a_{14, 15}:  &
& \big[\, (x_2 + 1) - x_2^2 \,\big]\, (z_2 + 1)^2\, \kappa^2 - 2\, (x_2 + 1)\, (z_2 + 1)\, \kappa + (x_2 + 1) = 0  &
\\
& a_{16, 17}:  &
& (x_2 + 1)\, (z_2 + 1)^2\, \kappa^2 - 2\, (x_2 + 1)\, (z_2 + 1)\, \kappa + \big[\, (x_2 + 1) - x_2^2 \,\big] = 0  &
\nonumber \\
& a_{18, 19}:  &
&  y\, (x_2 + 1)\, (z_2 + 1)^2\, \kappa^2 - 2\, (x_2 + 1)\, (z_2 + 1)\, \kappa + \big[\, y\, (x_2 + 1) + x_2^2\, (z_2 + 1) \,\big] = 0  &
\nonumber \\
& a_{20, 21}:  &
& y_2\, (z_2 + 1)\, \big[\, (x_2 + 1)\, (z_2 + 1) - x_2\, y_2 \,\big]\, \kappa^2 - 2\, y_2\, (x_2 + 1)\, (z_2 + 1)\, \kappa  &
\nonumber \\
&&
& + (x_2 + 1)\, \big[\, x_2\, (z_2 + 1) + y_2 \,\big] = 0  &
\nonumber
\end{align}
All $45$ canonical MIs of $\mathcal{T}_{2\digamma}$ can be expressed in terms of GPLs up to the order of $\epsilon^4$, except for $g_{36}$ at $\mathcal{O}(\epsilon^3)$ and $g_{36, 37, 38, 44, 45}$ at $\mathcal{O}(\epsilon^4)$, which are represented as one-fold integrals over GPLs. For illustration purposes, we further provide the third-order contribution to $g_{36}$ as follows:
\begin{align}
g_{36}^{(3)} = {} &
                    \smash{\int}
                    d\log \omega_{24}\,
                    \big[\,
                    4\, G(a_{1}, a_{7}; \kappa)
                    - 2\, G(a_{1}, a_{8}; \kappa)
                    - 2\, G(a_{1}, a_{9}; \kappa)
                    - 4\, G(a_{8}, a_{7}; \kappa)
\nonumber \\
&
                    + 2\, G(a_{8}, a_{8}; \kappa)
                    + 2\, G(a_{8}, a_{9}; \kappa)
                    - 4\, G(a_{9}, a_{7}; \kappa)
                    + 2\, G(a_{9}, a_{8}; \kappa)
                    + 2\, G(a_{9}, a_{9}; \kappa)
\nonumber \\
&
                    - 4\, G(a_{10}, a_{7}; \kappa)
                    + 2\, G(a_{10}, a_{8}; \kappa)
                    + 2\, G(a_{10}, a_{9}; \kappa)
                    - 4\, G(a_{11}, a_{7}; \kappa)
                    + 2\, G(a_{11}, a_{8}; \kappa)
\nonumber \\
&
                    + 2\, G(a_{11}, a_{9}; \kappa)
                    + 4\, G(a_{20}, a_{7}; \kappa)
                    - 2\, G(a_{20}, a_{8}; \kappa)
                    - 2\, G(a_{20}, a_{9}; \kappa)
                    + 4\, G(a_{21}, a_{7}; \kappa)
\nonumber \\
&
                    - 2\, G(a_{21}, a_{8}; \kappa)
                    - 2\, G(a_{21}, a_{9}; \kappa)
\nonumber \\
&
                    + 4\, G(a_{1}, a_{10}; \kappa)
                    + 4\, G(a_{1}, a_{11}; \kappa)
                    - 4\, G(a_{1}, a_{12}; \kappa)
                    - 4\, G(a_{1}, a_{13}; \kappa)
\nonumber \\
&
                    - 4\, G(a_{7}, a_{10}; \kappa)
                    - 4\, G(a_{7}, a_{11}; \kappa)
                    + 4\, G(a_{7}, a_{12}; \kappa)
                    + 4\, G(a_{7}, a_{13}; \kappa)
\nonumber \\
&
                    - 4\, G(a_{10}, a_{10}; \kappa)
                    - 4\, G(a_{10}, a_{11}; \kappa)
                    + 4\, G(a_{10}, a_{13}; \kappa)
                    + 4\, G(a_{10}, a_{12}; \kappa)
\nonumber \\
&
                    - 4\, G(a_{11}, a_{10}; \kappa)
                    - 4\, G(a_{11}, a_{11}; \kappa)
                    + 4\, G(a_{11}, a_{12}; \kappa)
                    + 4\, G(a_{11}, a_{13}; \kappa)
\nonumber \\
&
                    + 2\, G(a_{20}, a_{10}; \kappa)
                    + 2\, G(a_{20}, a_{11}; \kappa)
                    - 2\, G(a_{20}, a_{12}; \kappa)
                    - 2\, G(a_{20}, a_{13}; \kappa)
\nonumber \\
&
                    + 2\, G(a_{21}, a_{10}; \kappa)
                    + 2\, G(a_{21}, a_{11}; \kappa)
                    - 2\, G(a_{21}, a_{12}; \kappa)
                    - 2\, G(a_{21}, a_{13}; \kappa)
\nonumber \\
&
                    + \pi^{2}/3
                     \,\big]
\nonumber \\
 &
                    + d\log \omega_{25}\,
                    \big[\,
                    2\, G(a_{10}, a_{7}; \kappa)
                    - G(a_{10}, a_{8}; \kappa)
                    - G(a_{10}, a_{9}; \kappa)
                    + 2\, G(a_{11}, a_{7}; \kappa)
\nonumber \\
&
                    - G(a_{11}, a_{8}; \kappa)
                    - G(a_{11}, a_{9}; \kappa)
                    - 2\, G(a_{20}, a_{7}; \kappa)
                    + G(a_{20}, a_{8}; \kappa)
                    + G(a_{20}, a_{9}; \kappa)
\nonumber \\
&
                    - 2\, G(a_{21}, a_{7}; \kappa)
                    + G(a_{21}, a_{8}; \kappa)
                    + G(a_{21}, a_{9}; \kappa)
\nonumber \\
&
                    - G(a_{1}, a_{10}; \kappa)
                    - G(a_{1}, a_{11}; \kappa)
                    + G(a_{1}, a_{12}; \kappa)
                    + G(a_{1}, a_{13}; \kappa)
\nonumber \\
&
                    + 2\, G(a_{7}, a_{10}; \kappa)
                    + 2\, G(a_{7}, a_{11}; \kappa)
                    - 2\, G(a_{7}, a_{12}; \kappa)
                    - 2\, G(a_{7}, a_{13}; \kappa)
\nonumber \\
&
                    + G(a_{10}, a_{10}; \kappa)
                    + G(a_{10}, a_{11}; \kappa)
                    - G(a_{10}, a_{12}; \kappa)
                    - G(a_{10}, a_{13}; \kappa)
\nonumber \\
&
                    + G(a_{11}, a_{10}; \kappa)
                    + G(a_{11}, a_{11}; \kappa)
                    - G(a_{11}, a_{12}; \kappa)
                    - G(a_{11}, a_{13}; \kappa)
\nonumber \\
&
                    - G(a_{20}, a_{10}; \kappa)
                    - G(a_{20}, a_{11}; \kappa)
                    + G(a_{20}, a_{12}; \kappa)
                    + G(a_{20}, a_{13}; \kappa)
\nonumber \\
&
                    - G(a_{21}, a_{10}; \kappa)
                    - G(a_{21}, a_{11}; \kappa)
                    + G(a_{21}, a_{12}; \kappa)
                    + G(a_{21}, a_{13}; \kappa)
\nonumber \\
&
                    + G(a_{2}, a_{2}; x_{2})/2
                     \,\big]
\nonumber \\
&
                    - d\log \omega_{27}\,
                    \big[\,
                    2\, G(a_{1}, a_{2}; x_{2})
                    - G(a_{2}, a_{2}; x_{2})
                    \,\big]
\nonumber \\
&
                    - d\log (\omega_{28}/\omega_{34})\,
                    \big[\,
                    2\, G(a_{1}, a_{7}; \kappa)
                    - G(a_{1}, a_{8}; \kappa)
                    - G(a_{1}, a_{9}; \kappa)
\nonumber \\
&
                    - G(a_{2}; x_{2})\, G(a_{8}; \kappa)
                    + G(a_{2}; x_{2})\, G(a_{9}; \kappa)
                    - G(a_{2}, a_{2}; x_{2})
                     \,\big]
\nonumber \\
&
                    - d\log (\omega_{32}\, \omega_{35})\,
                    \big[\,
                    2\, G(a_{1}, a_{10}; \kappa)
                    + 2\, G(a_{1}, a_{11}; \kappa)
                    - 2\, G(a_{1}, a_{12}; \kappa)
                    - 2\, G(a_{1}, a_{13}; \kappa)
\nonumber \\
&
                    - G(a_{2}; x_{2})\, G(a_{10}; \kappa)
                    + G(a_{2}; x_{2})\, G(a_{11}; \kappa)
                    + G(a_{2}, a_{2}; x_{2})/2
                     \,\big]
\end{align}

\subsection{Linear basis of $\mathcal{T}_{3\digamma}$}
\label{appendix:B3}
The linear basis $\bm{g} = (g_1, \ldots, g_{15})^{T}$ of the top-branch $\mathcal{T}_{3\digamma}$ is presented in an iterative form. The expressions for all three MIs of the elliptic sector up to $\mathcal{O}(\epsilon^3)$ are given as follows:
\begin{equation}
g_{13} = \epsilon^3\, g_{13}^{(3)}\,,
\qquad\quad
g_{14} = \epsilon^3\, g_{14}^{(3)}\,,
\qquad\quad
g_{15} = \epsilon^2\, g_{15}^{(2)} + \epsilon^3\, g_{15}^{(3)}\,,
\end{equation}
with

where $\text{Cl}_{2}(x)$ denotes the Clausen function and $\text{Li}_{2}(x)$ represents the polylogarithm, both of order two, and the $J$ functions are defined by
\begin{equation}
J(f_1, f_2, \ldots, f_n)
=
\int_{0}^{1} d \kappa_1\, f_1(\kappa_1)
\int_{0}^{\kappa_1} d \kappa_2\, f_2(\kappa_2)
\; \cdots\,
\int_{0}^{\kappa_{n-1}} d \kappa_n\, f_n(\kappa_n)\,.
\end{equation}
The weights of the involved GPLs are specified as follows:
\begin{equation}
a_1 = 0\,,
\qquad
a_{2,3} = \pm\, 1\,,
\qquad
a_{4,5} = \pm\, i\,,
\qquad
a_{6,7} = \pm\, e^{i \pi/6}\,,
\qquad
a_{8,9} = \pm\, e^{-i \pi/6}\,.
\end{equation}
By differentiating the linear basis of MIs with respect to the integration path parameter $\kappa$, we obtain the following one-variable linear-form differential equation for $T_{3\digamma}$:
\begin{equation}
\frac{d}{d\kappa}\, \bm{g}(\bm{x}; \epsilon)
=
\Big[\,
\mathbb{A}_{\kappa}^{(0)}(\bm{x}) + \epsilon\, \mathbb{A}_{\kappa}^{(1)}(\bm{x})
\,\Big]\,
\bm{g}(\bm{x}; \epsilon)\,.
\end{equation}
The variables of the $J$ functions appearing in eq. \eqref{eq:Exp-branch3}, $\alpha_{i,j}$ and $\beta_{i,j}$, are matrix elements of the coefficient matrices $\mathbb{A}^{(1)}_{\kappa}$ and $\mathbb{A}^{(0)}_{\kappa}$, respectively.

\bibliographystyle{JHEP}
\bibliography{references}

\end{document}